\documentclass[12pt]{elsarticle}
\usepackage{amsmath}
\usepackage{xcolor}
\usepackage{stackengine}
\usepackage{braket}
\usepackage{bbold}
\usepackage{subcaption}
\usepackage{graphicx}
\usepackage[a4paper, total={7.5in, 10in}]{geometry}
\title{Simulations of distributed-phase-reference quantum key distribution protocols}
 \author{Venkat Abhignan, Abhishek Jamunkar, Gokul Nair, Mohit Mittal and Megha Shrivastava}
 \address{Qdit Labs Pvt. Ltd., Bengaluru - 560092, India}

\begin{document}
\makeatletter
\def\ps@pprintTitle{%
  \let\@oddhead\@empty
  \let\@evenhead\@empty
  \let\@oddfoot\@empty
  \let\@evenfoot\@oddfoot
}
\makeatother
\begin{frontmatter}
\begin{abstract}
    Quantum technology can enable secure communication for cryptography purposes using quantum key distribution. Quantum key distribution protocols provide a secret key between two users with security guaranteed by the laws of quantum mechanics. To define the proper implementation of a quantum key distribution system using a particular cryptography protocol, it is crucial to critically and meticulously assess the device's performance due to technological limitations in the components used. We perform simulations on the ANSYS Interconnect platform to characterise the practical implementation of these devices using distributed-phase-reference protocols differential-phase-shift and coherent-one-way quantum key distribution. Further, we briefly describe and simulate some possible eavesdropping attempts, backflash attack, trojan-horse attack and detector-blinding attack exploiting the device imperfections. 
\end{abstract}
\end{frontmatter}
\section{Introduction}

The inception of quantum key distribution (QKD) \cite{PhysRevLett.67.661} in quantum communication and cryptography has drawn a lot of attention because it offers two distant users Alice and Bob a secret key with composable and unconditional security provided by the principles of quantum mechanics \cite{RevModPhys.81.1301,Lo2014,RevModPhys.92.025002,Pirandola:20}. Beyond the fundamentally interesting Bennett-Brassard 1984 QKD protocol \cite{BENNETT20147}, distributed-phase-reference protocols such as differential-phase-shift (DPS) QKD \cite{PhysRevLett.89.037902} and coherent-one-way (COW) QKD \cite{Damien} have had significant experimental progress due to their relatively simpler setup. Both COW QKD \cite{Stucki:09,Stucki_2009,Walenta_2014,Korzh2015,Sibson2017,Sibson:17,roberts2017,Dai:20} and DPS DKD \cite{Takesue_2005,Diamanti:06,Takesue2007,Sasaki:11} have had possible variations in their implementation to improve their practicality and increase the secure distance of the secret key between the two users.

The typical components required to construct a QKD setup include single-photon sources, single-photon avalanche photo-diodes (SPADs), and other optical devices (such as beamsplitters, interferometers, etc). Due to inherent flaws in these components, the operation of a QKD protocol in practise differs from the ideal protocol and could let in eavesdroppers through several different backdoors \cite{IAQKD}. A common and effective method for understanding the dynamics of QKD systems and the protocol is to employ modelling and simulations \cite{Shuang2008,Buhari2012,mailloux2015,mailloux2016,Coles2016,PhysRevApplied.14.024036,Fan-Yuan2020,Anuj2022}. The theory-practice inconsistencies can also be improved by initially performing simulations. 

The first step of quantum hacking \cite{RevModPhys.81.1301,RevModPhys.92.025002,Jain2015} is to experimentally demonstrate the existence of the flaw in the QKD implementation in the presence of a probable deviation, such as one caused by the counterproductive behaviour of a particular set of components. Then, the impact of the loophole can be quantified, for instance, by performing simulations of the attack. To address this, we initially simulate distributed-phase-reference protocols DPS QKD and COW QKD on the ANSYS Interconnect platform in Sec. 2 and 3, respectively. Further, we perform some experimentally well-studied hacking attempts backflash attack \cite{Pinheiro:18,Meda2017}, Trojan-horse attack \cite{jmo2001,th2006,Jain_2014,Sajeed2017} and detector-control attack \cite{Lydersen2010,Wu:20} on these protocols considering some simple assumptions in Sec. 4, 5 and 6, respectively.

\section{Simulation of DPS protocol}
The DPS protocol \cite{PhysRevLett.89.037902} includes an Alice module, which creates coherent pulses from the continuous laser source using an intensity modulator. The coherent pulse phase is randomly modified using a phase modulator and random number generator. Alice applies a phase shift of 0 or $\pi$ based on whether the random number generator output is 0 or 1, and the bit information is encoded in the difference between the phases of successive pulses. After receiving the optical pulses from Alice through a quantum channel, Bob routes them via a delay line interferometer, which has two output ports that are linked to single photon detectors. When the phase difference between two successive pulses is zero, detector D1 clicks; when the phase difference is $\pi$, detector D2 clicks. Bob keeps track of which detector clicks in which time frame, related to Alice and Bob's synchronised clock. Then Bob shares the measured photon's arrival time with Alice over a previously defined classically authenticated channel. Alice, on her side, deletes data from the instances when Bob fails to find a photon. As a result, Bob and Alice have created a sifted key. 

\begin{figure}[hp]
\centering
\includegraphics[width=1\columnwidth]{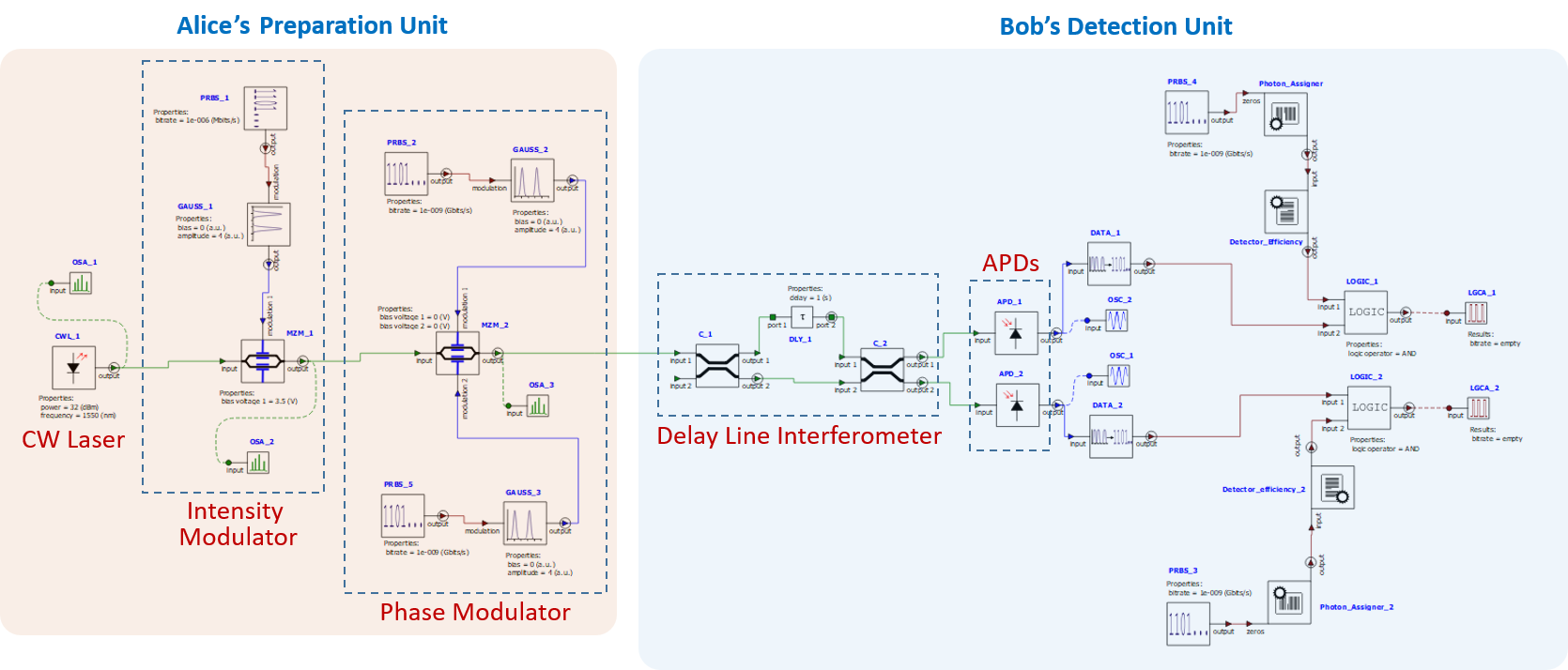}
\caption{ANSYS Interconnect simulation for DPS protocol}
\end{figure}

The image in Fig. 1 shows a DPS protocol simulation on the ANSYS Interconnect platform. A continuous optical source is modelled using a single-mode continuous wave (CW) laser element from the ANSYS Interconnect library. CW laser generates a single-frequency optical signal at a constant amplitude. The phase and polarization of the laser are well-defined and do not vary during the simulation. An intensity modulator is used to transform a CW optical signal into a train of optical pulses.

A Mach-Zehnder modulator (MZM) component from the ANSYS Interconnect library is used to model the intensity and phase modulators. In the MZM, optical input ($E_{in}$) is split into two arms (upper and lower) and subsequently, phase modulated using phase shifts $\phi_1$ and $\phi_2$ driven by the electrical signals $V1$ and $V2$. The two modulated signals are then recombined to give optical output $E_{out}$, given by \begin{equation}
    E_{out} = E_{in} \left(\text{e}^{i\phi_1}+\text{e}^{i\phi_2}\right)
\end{equation} 
where phase shifts $\phi_1$ and $\phi_2$ are
\begin{equation}
    \phi_{1,2}=\pi\left(\frac{V^{1,2}}{V^{\pi}_{\text{RF}}}+\frac{V_{\text{bias}}^{1,2}}{V^{\pi}_{\text{DC}}}\right).
\end{equation}
Here $V^{1,2}$ and $V^{1,2}_{\text{bias}}$ are the input RF voltage and bias voltage applied to two arms with RF voltage $V^{\pi}_{\text{RF}}$ and DC voltage $V^{\pi}_{\text{DC}}$. 

The CW laser source is used as an optical signal input, which will be modulated by an intensity modulator by the radio frequency (RF) voltage from the pulse generator. The pulse generator driven by pseudo-random binary sequence (PRBS) produces periodic electrical pulses (by setting PRBS\_1 output to “1”s). The MZM (MZM\_1) is operated in the “balanced single drive” mode, where the equal and opposite voltages (from the RF pulse generator) are applied to the two arms. The CW laser source is transformed into the pulsed signal using the appropriate bias voltage. 

Further, the optical pulses from the intensity modulator are then passed through a phase modulator. The phase modulator is realized using an MZM (MZM\_2) with the same RF voltage applied to both arms. This results in optical pulses with phase randomly modulated by $0 (\pi)$ depending on the PRBS (PRBS\_2 and PRBS\_5 with same output) input $0(1)$ to the RF pulse generator. These optical pulses, randomly phase-modulated by $0$ or $\pi$, are transmitted to Bob. 

In the simulation, we do not attenuate the optical signal, and hence, all the pulses are received at Bob’s delay line interferometer (DLI). To realize a DLI in the ANSYS Lumerical Interconnect simulation tool, we used two $2\times2$ coupler ($50:50$ coupling ratio) with an optical time delay in one path (C\_1 and C\_2 with DLY\_1)to retrieve the phase-encoded by Alice. The optical time delay equals the separation between two consecutive optical pulses. The signal from two output ports of the second coupler is fed to photodetectors. Our simulation used the Avalanche photodetector (APD) element of ANSYS Lumerical Interconnect (APD\_1 and APD\_2). The output of this simulation is further analyzed in simulations with attacks.

\section{Simulation of COW protocol}
The COW protocol \cite{Damien} includes an Alice that generates a series of optical states, $0_t\alpha_{t-\tau}$, $\alpha_t 0_{t-\tau}$, and $\alpha_t\alpha_{t-\tau}$, using an intensity modulator and a laser source where $\alpha$ is mean photon number of the optical pulse. She sends these states to Bob, with $0_t$ usually representing the vacuum state or no pulse and a bit value at time $t$. These signals represent a bit value of 0, a bit value of 1, and a decoy signal in that order. The incoming signals are separated into the data and monitoring lines at Bob's end by a beam splitter with a specific transmittance. The data line uses a detector ($D_B$) to measure the arrival time of each signal to distinguish between the bit states $0_t\alpha_{t-\tau}$ and $\alpha_t 0_{t-\tau}$. To detect potential eavesdropping, the monitoring line examines the coherence of neighbouring non-empty pulses. A Mach-Zehnder interferometer with an imbalance of $\tau$ is used for this examination, and two detectors, $D_{M1}$ and $D_{M2}$, are positioned after it. The configuration of the interferometer prevents neighbouring coherent states $\alpha$ from causing a detection in one of the detector $D_{M1}$ or $D_{M2}$ (based on our choice in delay line interferometer) and a detection in another detector. Visibilities $V_s$ are used to quantify errors in the monitoring line defined by \begin{equation}
    V_s = \left|\frac{\hbox{Detection} (D_{M1})-\hbox{Detection} (D_{M2})}{\hbox{Detection} (D_{M1})+\hbox{Detection} (D_{M2})}\right| 
\end{equation} with $s \in \{``d",``01",``0d",``d1",``dd"\}$. $s=``d"$ and $s=``01"$ correspond to decoy signal and signals for bit 1 followed by bit 0, respectively. The other sequences in $s$ are defined similarly. Detection ($D_{M1}$) here means that there is a detection in $D_{M1}$ detector with an electrical output beyond a threshold. When Alice releases a bit state of 0 or 1, Bob detects them after classically communicating with Alice the time frame of detections and positions of decoy states to form the sifted key. They then extract a final secure key from the sifted key via privacy amplification. 

The image in Fig. 2 shows a COW simulation on the ANSYS Interconnect platform. Compared to earlier Fig. 1 of DPS simulation, Alice's module in COW consists only of the intensity modulator on the left side of Fig. 1. Similar to DPS protocol simulation, a MZM (MZM\_1) component from ANSYS Interconnect library is used to model intensity modulator. The pulse generator produces an electrical pulse depending on the PRBS (PRBS\_1) input. The PRBS is set to make a random sequence of “$1-0$” (for logical 0), “$0-1$” (for logical 1), and “$1-1$” (for decoy). Similar to DPS simulation, we do not attenuate the optical signal; hence, all the pulses are received at Bob’s end.

Bob's module consists of a 90:10 splitter, which splits and transmits 90\% and 10\% of the photons that it receives to data and monitoring lines, correspondingly. ANSYS Interconnect does not offer a $90:10$ fibre coupler. Therefore, a custom element (COMPOUND\_1) is designed to achieve this ratio. Moreover, since the mean photon number $> 1$ in this idealistic simulation, all the pulses are detected in the data and monitoring lines. The data line consists of an APD ($D_B$), and the Monitoring line consists of a delay line interferometer with a pair of APDs ($D_{M1}$ and $D_{M2}$). For instance, when Alice sends bits of information 01d10001d1 with $t= 1s$ and $\tau=0.5s$, we can observe from Fig. 3 that APD\_1 ($D_{B}$ with electrical output in oscilloscope OSC\_3) gives the actual key information and APD\_2 ($D_{M1}$ with output in OSC\_1) detects when there are two consecutive pulses (all the pulses have intensities with arbitrary units). Further, from Eq. 1, the visibility $V_s=1$ since detection from APD\_3 ($D_{M2}$ with output in OSC\_2) is negligible due to destructive interference. 

A simple manner to check Eve's (An eavesdropper) presence in this protocol is when she interferes and disturbs coherence between neighbouring pulses. We introduce a phase modulator (similar to the phase modulator in DPS from the previous section)  in between Alice's and Bob's modules to assign phase $00000\pi\pi\pi\pi\pi\pi\pi\pi\pi\pi\pi\pi000$
as can be seen in Fig. 4 to observe a disturbance of coherence such that the output of the monitoring line gets modified, such as in Fig. 5. APD\_2 detects when the coherence is preserved, and APD\_3 detects the vice-versa with a $V_s=1/5$.

\begin{figure}[htp]
\centering
\includegraphics[width=1\columnwidth]{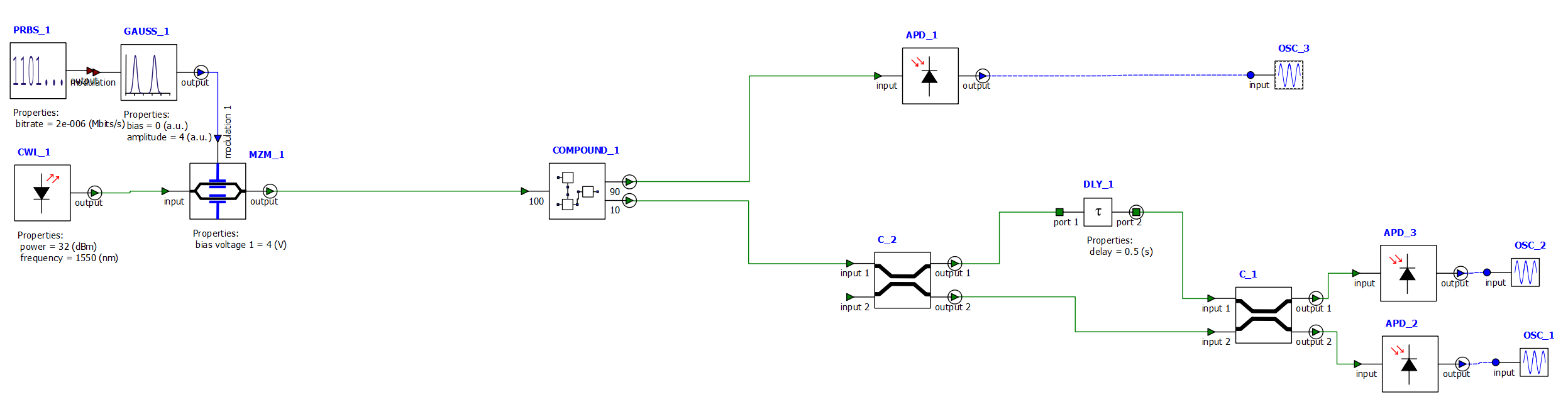}
\caption{ANSYS Interconnect simulation for COW protocol}
\end{figure}
\begin{figure}[htp]
\centering
\begin{subfigure}{0.45\linewidth}
\includegraphics[width=\linewidth]{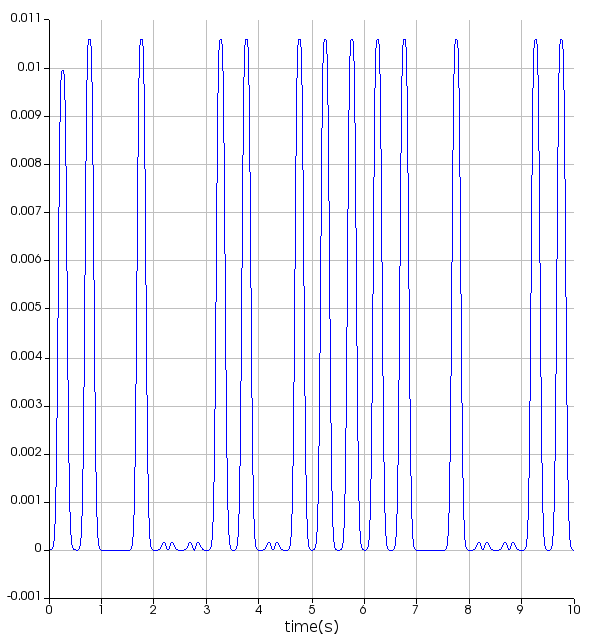}
\caption{Monitoring line, OSC\_2}
\end{subfigure}
\begin{subfigure}{0.45\linewidth}
\includegraphics[width=\linewidth]{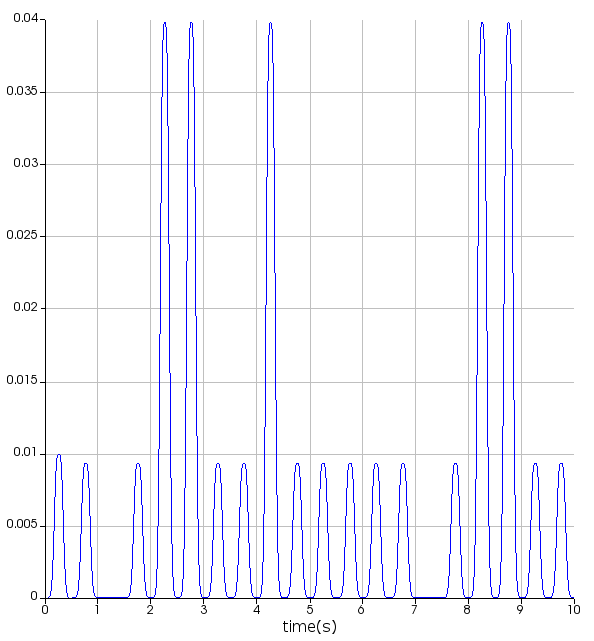}
\caption{Monitoring line, OSC\_1}
\end{subfigure}
\begin{subfigure}{0.45\linewidth}
\includegraphics[width=\linewidth]{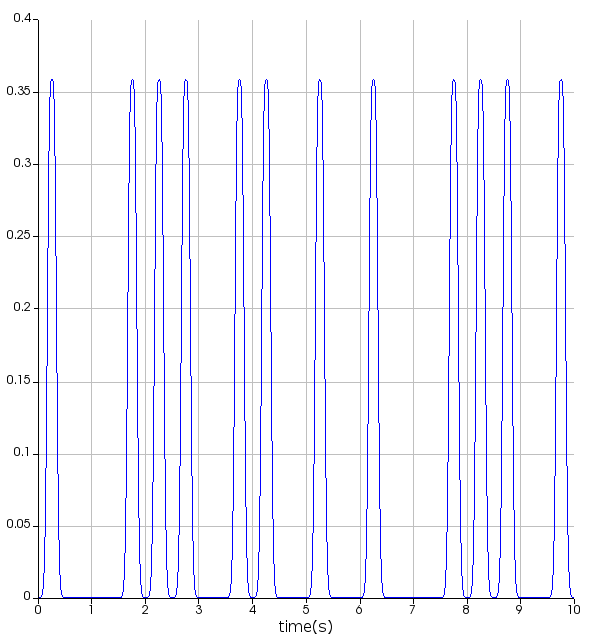}
\caption{Data line, OSC\_3}
\end{subfigure}
\caption{Comparing electrical output of monitoring line OSC\_2 and OSC\_1 (APD\_3 and APD\_2), data line OSC\_3 (APD\_1) in DPS.}
\end{figure}
\begin{figure}[htp]
\centering
\includegraphics[width=1\columnwidth]{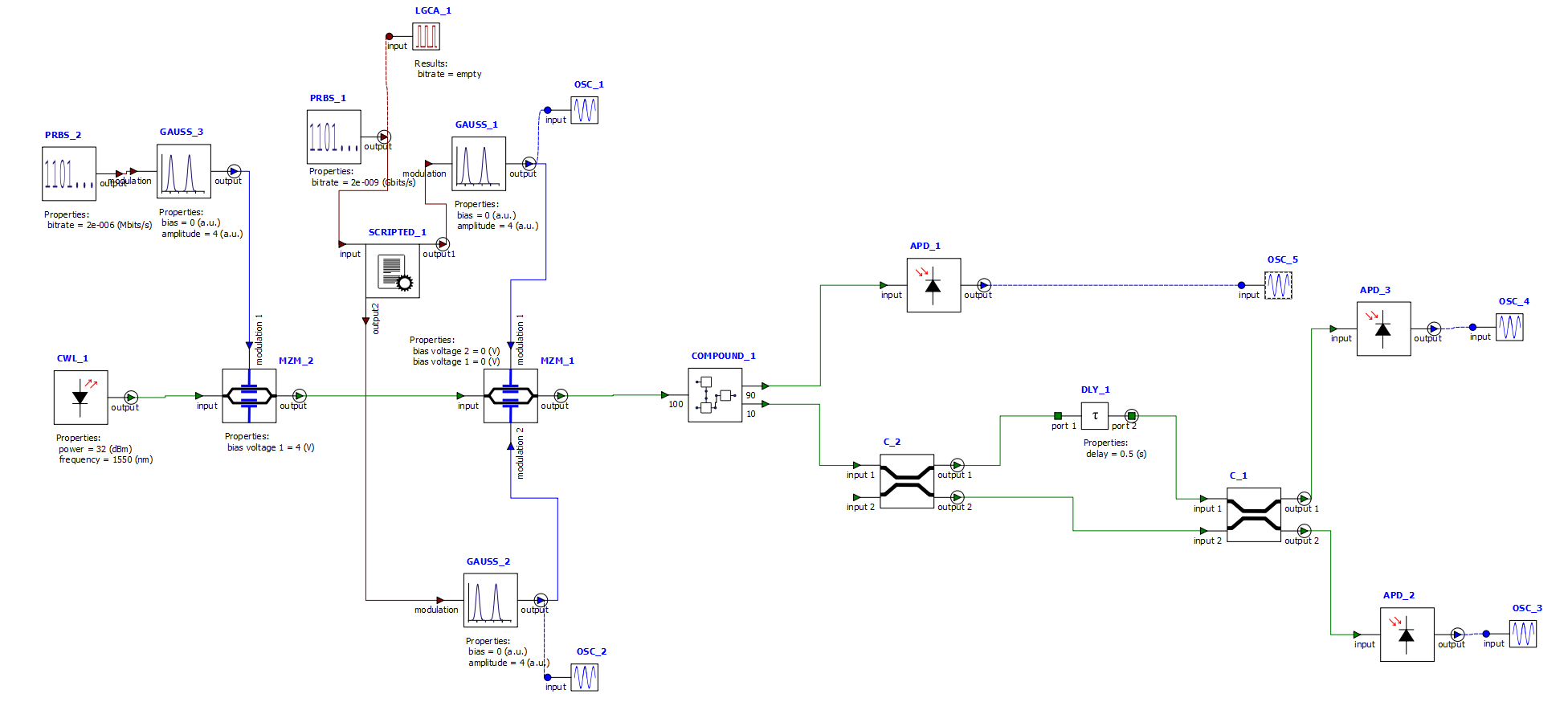}
\caption{ANSYS Interconnect simulation for COW protocol with non-unity in $V_s$.}
\end{figure}
\begin{figure}[htp]
\centering
\begin{subfigure}{0.45\linewidth}
\includegraphics[width=\linewidth]{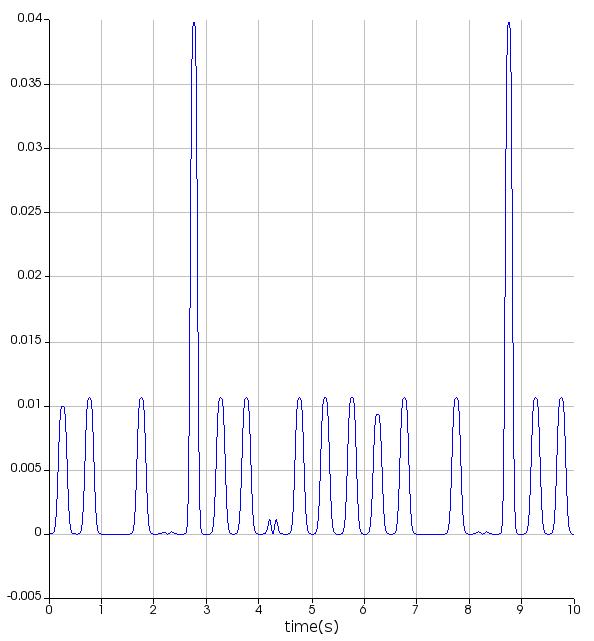}
\caption{Monitoring line, OSC\_4}
\end{subfigure}
\begin{subfigure}{0.45\linewidth}
\includegraphics[width=\linewidth]{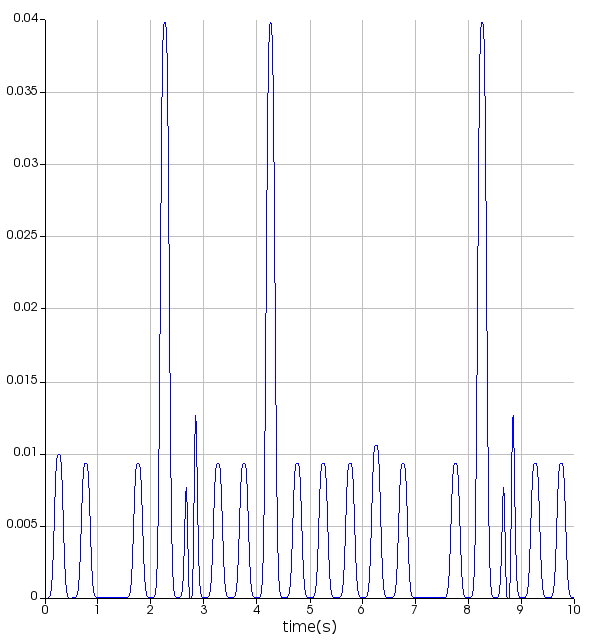}
\caption{Monitoring line, OSC\_3}
\end{subfigure}
\begin{subfigure}{0.45\linewidth}
\includegraphics[width=\linewidth]{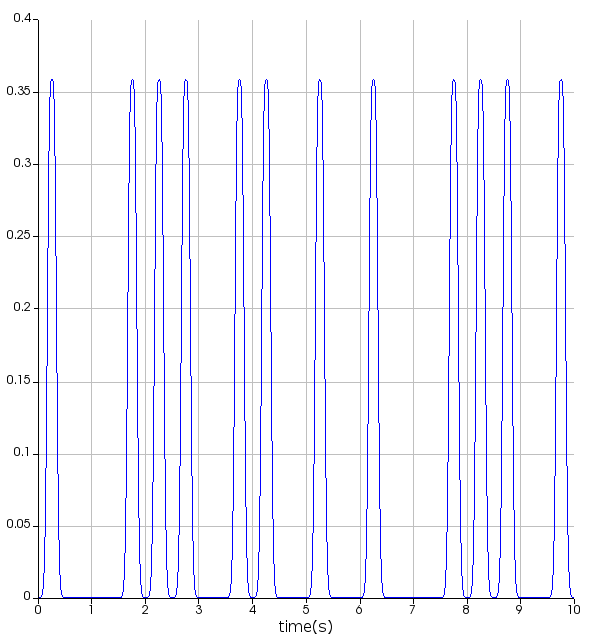}
\caption{Data line, OSC\_5}
\end{subfigure}
\caption{Comparing electrical output of data line OSC\_5 (APD\_1), monitoring line OSC\_4 and OSC\_3 (APD\_3 and APD\_2) in DPS.}
\end{figure}
\section{Backflash Attack}
\subsection{Introduction}
  Most commercially available QKD setups use SPADs on Bob's side receivers. SPADs are semiconductor tools enabling Bob to record information at the single-photon level. These are semiconductor devices that are similar to photodiodes and are capable of converting light into electrical current. Usually, the p-n junction of photodiodes, when exposed to photons, produces a current flow. Meanwhile, when SPADs are biased above their breakdown voltage, they function in the Geiger mode, a modified photodetector utilised in single photon detection. An avalanche breakdown is sparked when a photon hits the SPAD, producing a distinct avalanche current. This distinction is vital because photodiodes are not designed for the Geiger mode employed by SPADs; instead, they typically function under different voltage settings. Even though it has been discovered that SPADs can be used for eavesdropping attacks, most SPADs have a significant vulnerability that has gone unnoticed, called as backflash attacks \cite{Pinheiro:18,Meda2017}. The voltage to the p-n junction of the photodiode is much beyond the breakdown voltage given in the Geiger mode of SPADs, which is how they are intended to operate. In such a case, a single absorbed photon might cause the SPAD to cause self-sustaining discharge. The arrival time of the detected photon can then be determined with high timing accuracy, together with the avalanche current. The avalanche current must be extinguished to reset the SPAD and prepare it for subsequent detection (creating the detector's dead time). The quenching is carried out via a quenching circuit. Newman discovered that the photon absorption causing an avalanche of charge carriers in silicon could lead to photon emission causing the backflash effect \cite{PhysRev.100.700}. 

\subsection{Quantifying backflash attacks}
It was experimentally shown that there is a linear relationship between the avalanche current and backflash photon emission for an InGaAs SPAD \cite{6570518}. This approach has been proposed as a non-invasive way to estimate the avalanche current waveform of a SPAD, which encloses the shared information in case of QKD communication (Depending on the type of QKD protocol, whether information is stored in time bins (COW) or encoded in phase between time bins (DPS)). Backflash photon characteristics are primarily determined by how the SPAD is implemented in practice, the semiconductor being employed and specifically by the quenching electronic parameters. This explains the differences in the backflash spectral distribution, temporal profile, and intensity observed in various SPAD models operating in multiple regimes \cite{Meda2017}.

\subsection{ANSYS simulation of blackflash attack}

For implementing the backflash attack on the DPS ANSYS Interconnect simulation, we modify Bob's setup, such as in Fig. 6. Compared with Fig. 1, it can be observed that MOD\_APD\_1 and MOD\_APD\_2 are modified APDs. We modify APDs so that they emit back photons proportional to the photons they are incident with. We make this assumption from the deduction in Ref. \cite{6570518}, where avalanche current has characteristics of backflash photon. So, we consider every backflash photon to have an imprint of the original photon (including phase information). Further, we introduce Eve's module, such as in Fig. 7. It consists of an optical circulator such that information from Alice will reach Bob. In contrast, information from Bob will only reach Eve. Further, she will possess Bob's original setup (This knowledge is considered to be known to everyone) to detect the backflash photons. The entire design of the backflash attack on DPS comprising of Alice, Bob and Charlie can be seen in Fig. 8. Comparing the digital output of corresponding D1 detectors of Bob's module and Eve's module, we can observe that the entire key can be obtained in this ideal assumption as seen in Fig. 9 (Output from MOD\_APD\_1, APD\_2 in Fig.s 6, 7 are fed to digital to analog converters DATA\_1, DATA\_2 and are studied in LGCA\_5, LGCA\_10, respectively). This is not realistic since the average number of electrons that the APD under test produces in each avalanche is $2.7 \times 10^8$, and the chance of backflash photon production per avalanche electron is $2.4 \times 10^{-10}$ \cite{Pinheiro:18}, which has to be introduced further.

\begin{figure}[htp]
\centering
\includegraphics[width=1\columnwidth]{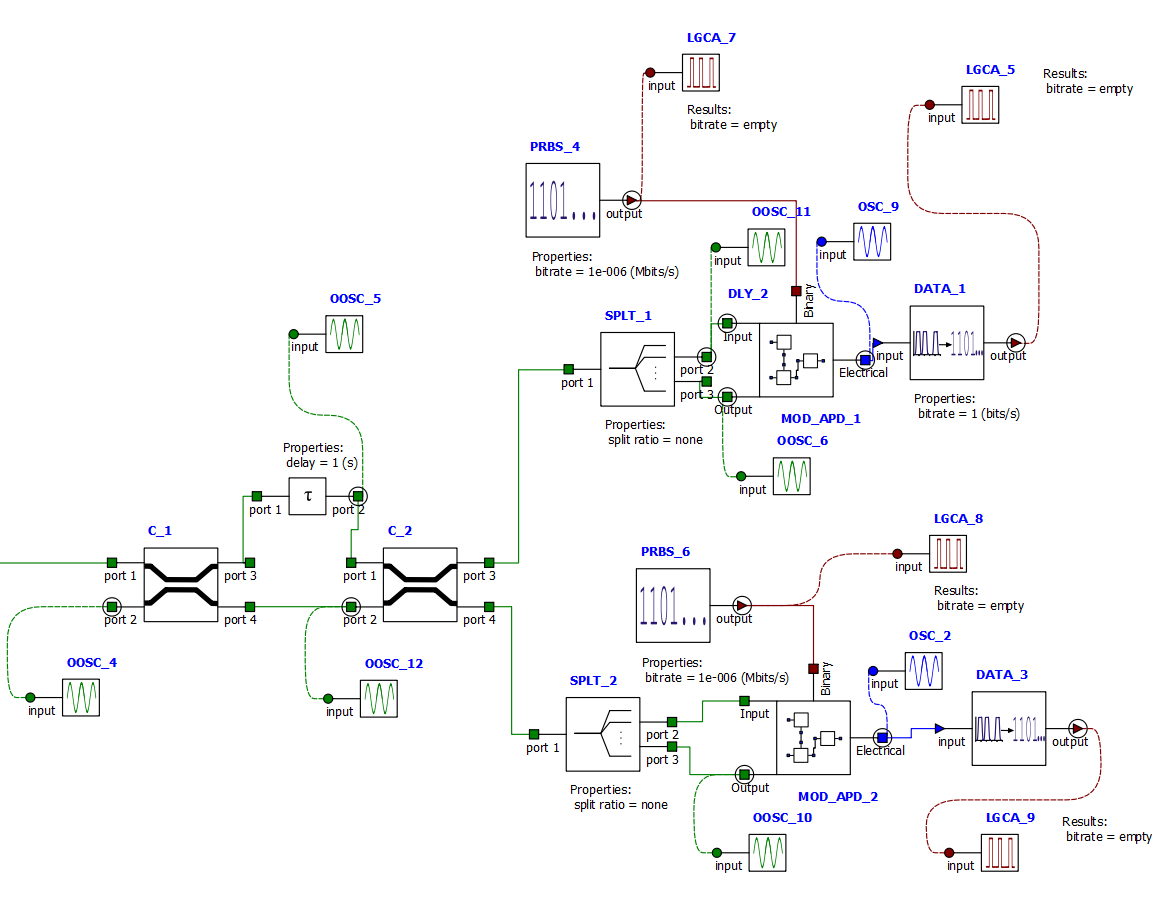}
\caption{Bobs module of ANSYS Interconnect simulation for backflash attack on DPS}
\end{figure}
\begin{figure}[htp]
\centering
\includegraphics[width=1\columnwidth]{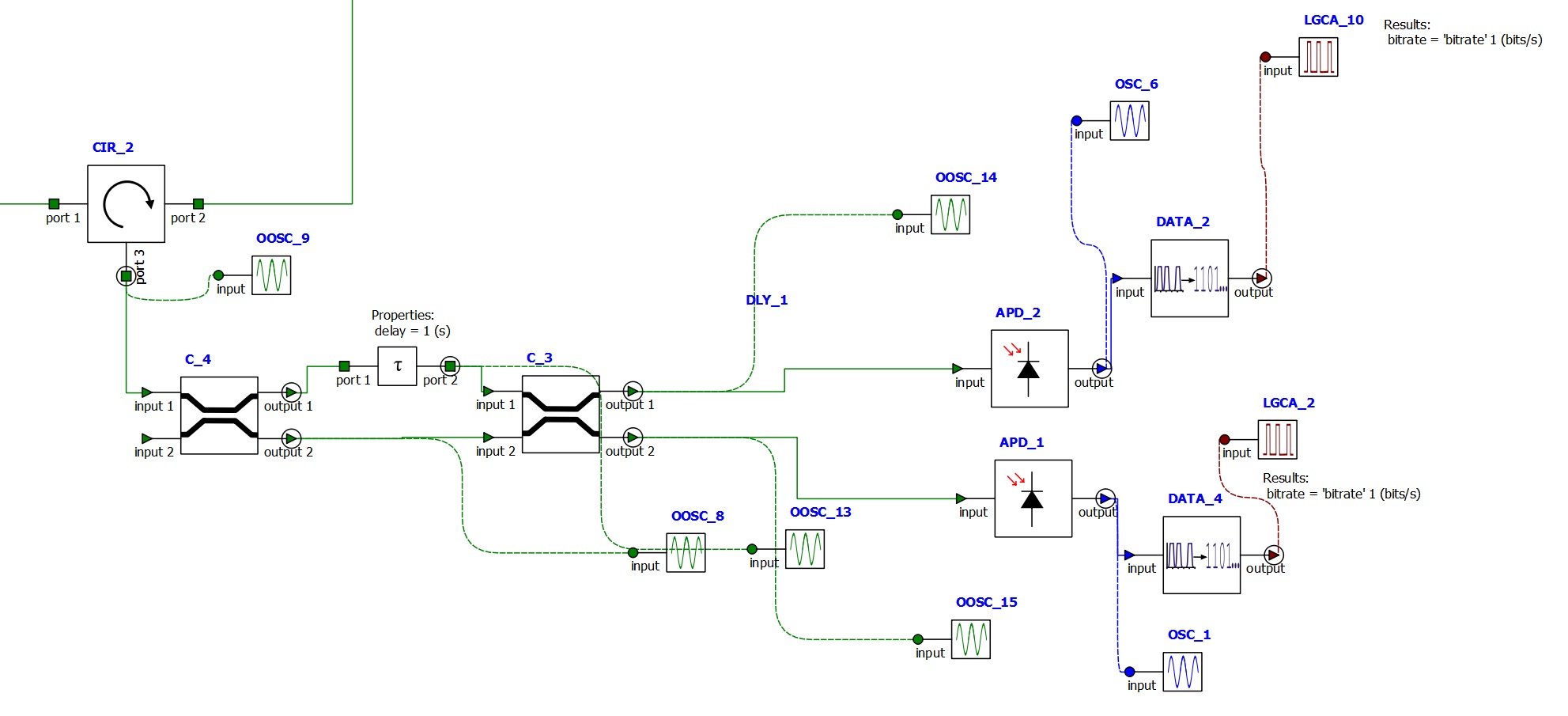}
\caption{Eves module of ANSYS Interconnect simulation for backflash attack on DPS}
\end{figure}
\begin{figure}[htp]
\centering
\includegraphics[width=1\columnwidth]{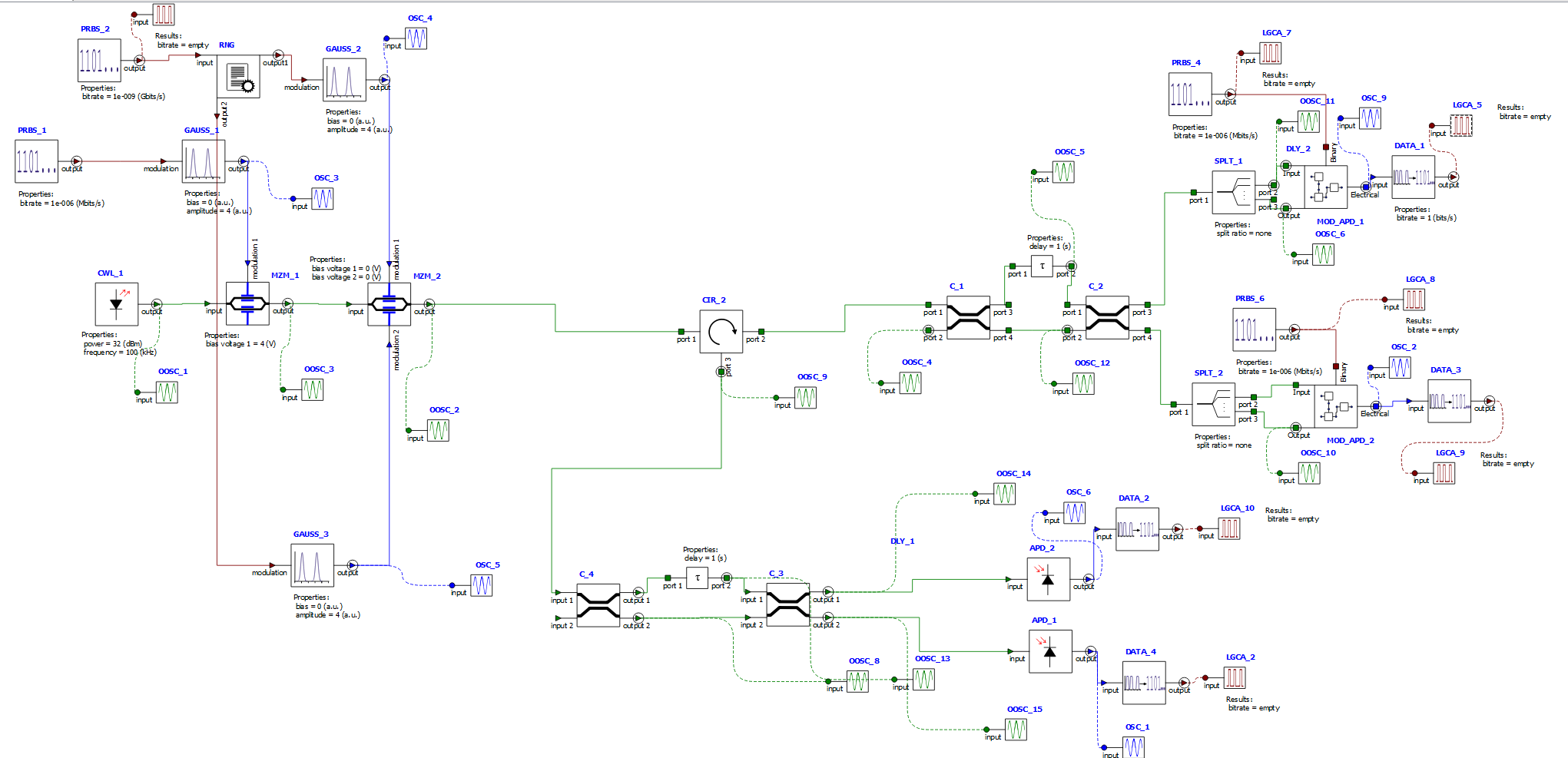}
\caption{ANSYS Interconnect simulation for backflash attack on DPS}
\end{figure}
\begin{figure}[htp]
\centering
\begin{subfigure}{0.4\linewidth}
\includegraphics[width=\linewidth]{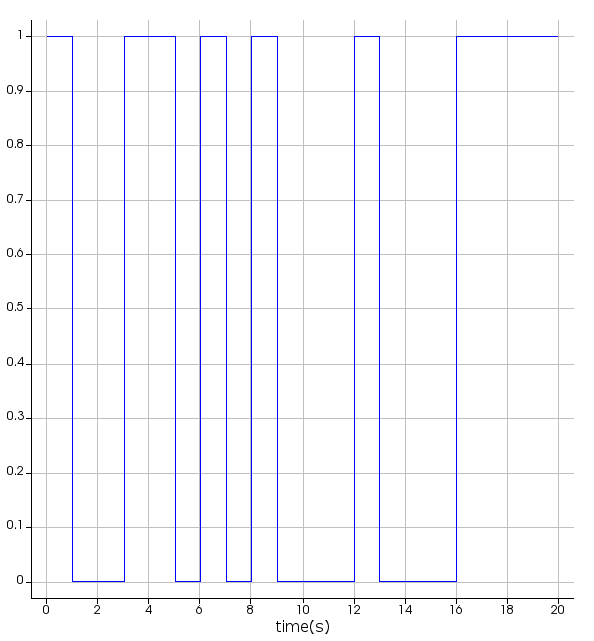}
\caption{D1 detector of Bob's module}
\end{subfigure}
\begin{subfigure}{0.4\linewidth}
\includegraphics[width=\linewidth]{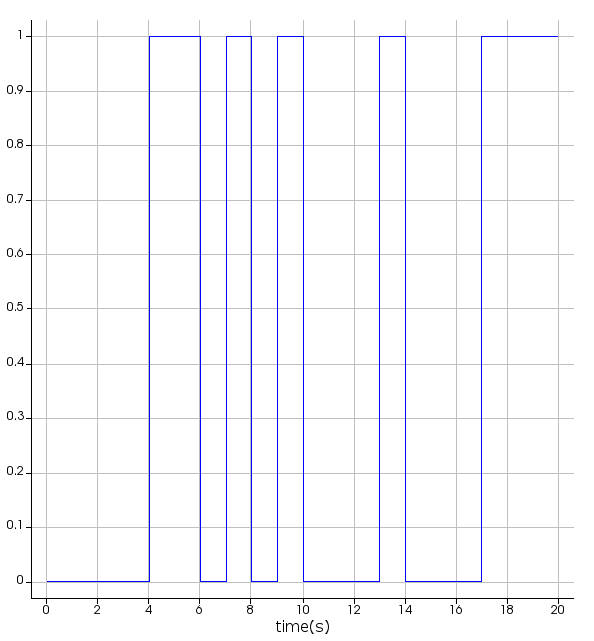}
\caption{D1 detector of Eve's module}
\end{subfigure}
\caption{Digital output of corresponding D1 detectors of Bob's module and Eve's module in DPS under backflash attack}
\end{figure}

For implementing the backflash attack on COW protocol, similar to ANSYS simulation of backflash on DPS protocol, we modify Bob's setup, such as in Fig. 10. Compared with Fig. 2, it can be observed that MOD\_APD, MOD\_APD\_1, and MOD\_APD\_2 are modified APDs that emit back photons. We can find similarities in the optical information by comparing the optical output of Bob's data line and Eve's optical circulator in Fig. 11. 

\begin{figure}[htp]
\centering
\includegraphics[width=1\columnwidth]{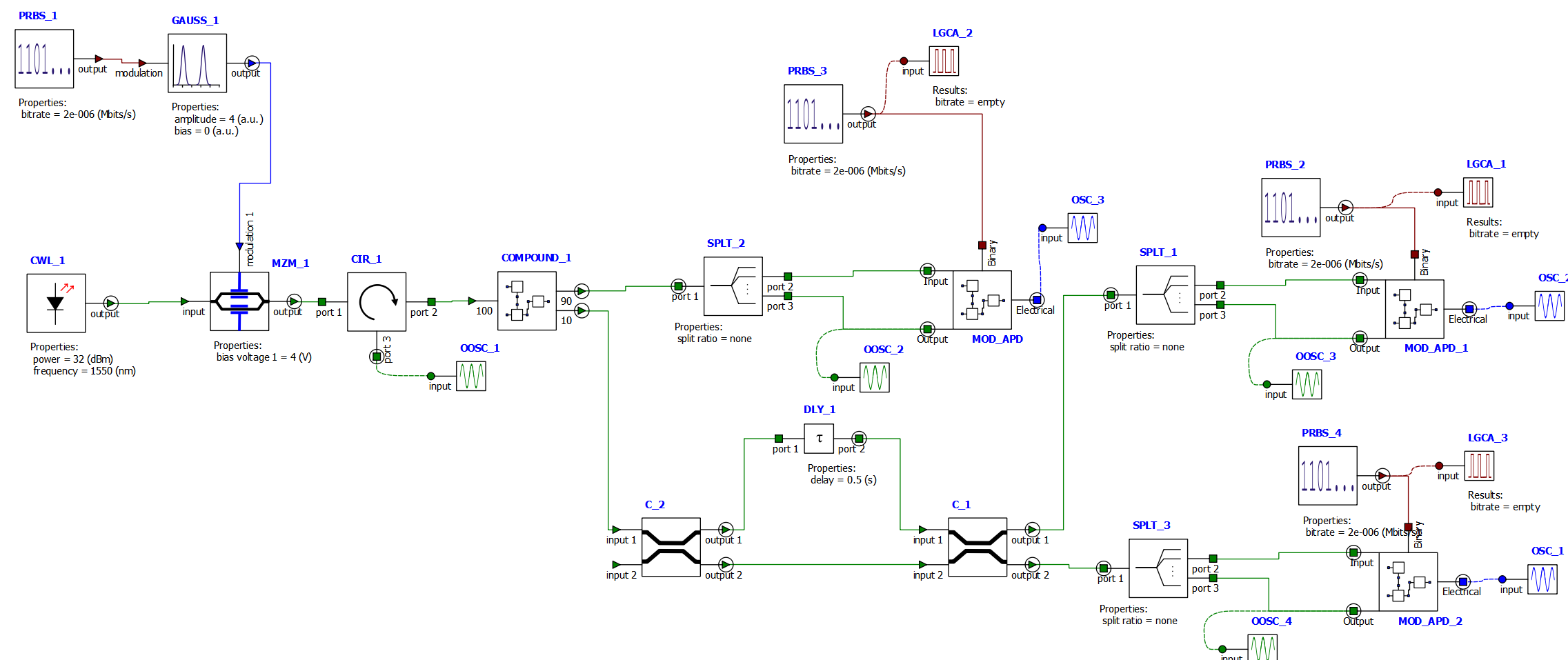}
\caption{ANSYS Interconnect simulation for backflash attack on COW protocol}
\end{figure}
\begin{figure}[htp]
\centering
\begin{subfigure}{0.45\linewidth}
\includegraphics[width=\linewidth]{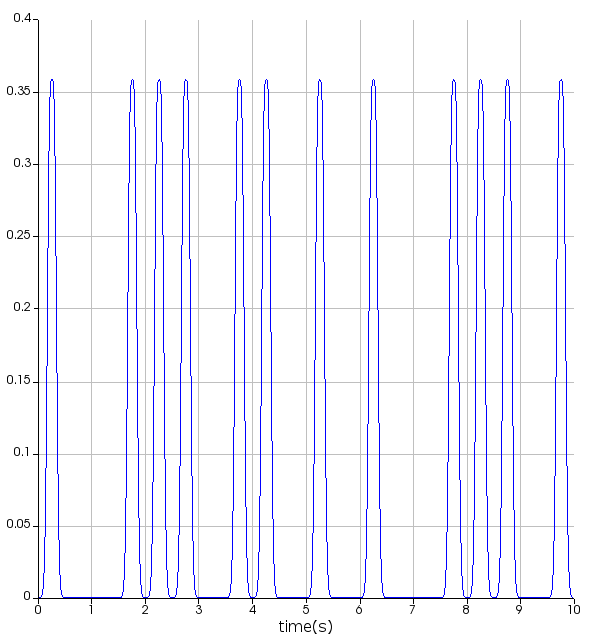}
\caption{Data line, OSC\_3}
\end{subfigure}
\begin{subfigure}{0.45\linewidth}
\includegraphics[width=\linewidth]{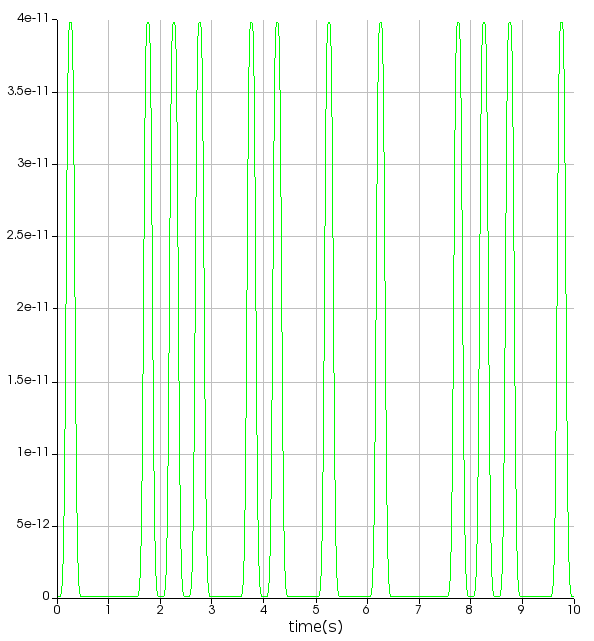}
\caption{Optical output from isolator of Eve}
\end{subfigure}
\caption{Output of data line of Bob's module and Eve's optical circulator.}
\end{figure}

\subsection{Countermeasures}

To prevent attacks based on backflash photons, QKD systems should be adequately designed and tested. Passive optical devices like circulators, isolators, or spectral filters may be used as potential solutions to stop backflashes from escaping a QKD system. These countermeasures should consider the broad emission bandwidth of backflash photons. 

\section{Trojan-horse attack}
\subsection{Introduction}
The name "Trojan-horse" is a misrepresentation in the context of practical QKD since most attacks in this category do not involve Alice or Bob accepting seemingly benign objects from Eve; this phrase has been taken from classical cryptography. The fundamental concept underlying these attacks is that light is reflected or scattered back as it passes through optical components \cite{jmo2001}. For instance, a Fresnel reflection happens when the refractive indices of two components change at an interface. A specular reflection is produced if the light strikes mirror-like surfaces. By using the quantum channel to transmit light pulses (trojan pulses) into Alice/Bob's equipment and scanning through the many reflections to gather the necessary data, Eve can actively lay the groundwork for such an attack \cite{th2006}. An optical isolator is one of the most well-known defences against Trojan-horse attacks for real-world QKD systems. When light comes from one end, this device allows it to pass, but when it comes from the other end, it prevents it \cite{Jain_2014}. However, Eve can select another wavelength where the isolator's extinction level is insufficient \cite{Sajeed2017}. 

\subsection{Practical implementations of Trojan-horse in commercial products}
As a proof-of-principle, Trojan-horse was recently demonstrated operating against the quantum cryptosystem Clavis2 from ID Quantique \cite{Jain_2014}. Installing a passive monitoring device would be a simple way for Alice to identify a Trojan-horse attack before it even occurs from a suitable detector. It monitors the incoming signal and alerts when pre-defined thresholds are crossed. This countermeasure, however, cannot be applied directly to the Bob subsystem because a passive monitoring device would further reduce the secret key rates by introducing undesired attenuation in the weak states of light coming from the quantum channel, which was exploited in this work. Eve could identify Bob's basis choice in the modified BB84 protocol (SARG04) of Clavis2 and, consequently, the raw key bit with more than 90\% probability using only a few back-reflected photons per cycle. However, Bob's single-photon detectors experience a considerable degree of after-pulsing due to Eve's intense pulses, which leads to a high quantum bit error rate (mismatch of reconciled bits between Alice and Bob) that effectively guards against this kind of attack. Ideally, the SPAD will return to a quiescent state after the avalanche occurrence, prepared to detect the next photon. On the other hand, after-pulsing is an undesirable consequence that may result in delayed or erroneous detection events. The cause of after-pulsing is the trapping of charge carriers (holes or electrons) in semiconductor material flaws in the SPAD during the avalanche phase. Later, these imprisoned charge carriers may be released, resulting in more avalanches and erroneous photon detection events. Usually, after-pulses appear as delayed, frequently haphazardly timed detection events that come after a real photon detection.

More recently, a nearly "invisible" Trojan-horse attack was performed on the same Clavis2 device at 1924 nm as that of 1550 nm performed by the previous work \cite{Sajeed2017}. They offered experimental proof that a Trojan-horse attack would be successful if Eve produced bright pulses at a wavelength where the SPADs after-pulsing is much reduced. The fundamental principle is that photons with energy below the bandgap of the material used in the SPAD absorption layer (InGaAs) primarily pass through the material unabsorbed, leading to very little afterpulsing. In other words, at comparatively longer wavelengths of 1924 nm, the SPAD exhibits significantly less afterpulsing than at 1550 nm. Nevertheless, the attenuation is typically more significant; for example, Eve would not just need to inject a higher mean photon number into Bob to receive a higher mean photon number in the back-reflection, as the optical loss through the 1924 nm Trojan-horse attack is 20 dB more than that at 1550 nm.

 \subsection{ANSYS simulation of Trojan Horse attack}

Primarily, since back reflection is used for performing this attack, Eve can send an intensity-modulated pulse from a different wavelength source (1000 nm) compared to Alice's wavelength source (1550 nm), and reflections from Alice's intensity modulator can be detected to get same phase modulation as performed for the information sent to Bob, provided the timing of the back-reflected pulse matches with the modulation of the phase modulator. As shown in Fig. 12, Eve can further use an optical filter to filter 1000 nm (OBP\_2 on the bottom right of Fig. 12) phase-modulated pulses to detect the key, whereas she can send the original 1550nm (OBP\_1 on the top right of Fig. 12) phase modulated pulses to Bob. Compared with the DPS simulation without hacking in Fig. 1, it can be seen that COMPOUND\_1 is the modified intensity modulator that sends back reflection. The additional SPLT\_2 (Optical splitter) is required since the Mach-Zehnder modulator in ANSYS Interconnect is not bidirectional, but Mach-Zehnder modulators are bidirectional in general. After filtering, the output of these two different optical pulses can be seen in Fig. 13; OSA\_1 (OBP\_1) and OSA\_2 (OBP\_2) are the optical outputs from optical spectrum analyzers. Eve will further employ the same setup as Bob, as seen in Fig. 14, to get the identical phase-modulated optical pulses in their corresponding D1 detectors. These pulses of corresponding D1 detectors are similar, as seen in Fig. 15. However, one has to realize that both the incoming Trojan-horse pulses and reflected pulses from Alice side get phase modulation at different time intervals depending on the distance of propagation of these pulses and surface of incidence which needs to be sorted using an optical time domain reflectometer (OTDR) \cite{Jain_2014,Sajeed2017}. Also, phase modulation at different wavelengths will generally be distinct. So, even if Eve deploys the same setup, she may not get the information from the secret key ideally. We ignore these in the simple simulation due to the complications that must be further considered.

A similar attack can be performed on COW protocol, as seen in Fig. 16, where we send only a laser signal from a different wavelength source instead of intensity-modulated pulses. Here, COMPOUND\_1 is the modified laser that sends back reflection. Beyond this ideal simulation, one has to measure the actual number of reflected photons after attenuation and reflections at multiple surfaces of Alice's side based on the transmission of optical components implemented to determine how much information can leak.  

   \begin{figure}[htp]
\centering
\includegraphics[width=1\columnwidth]{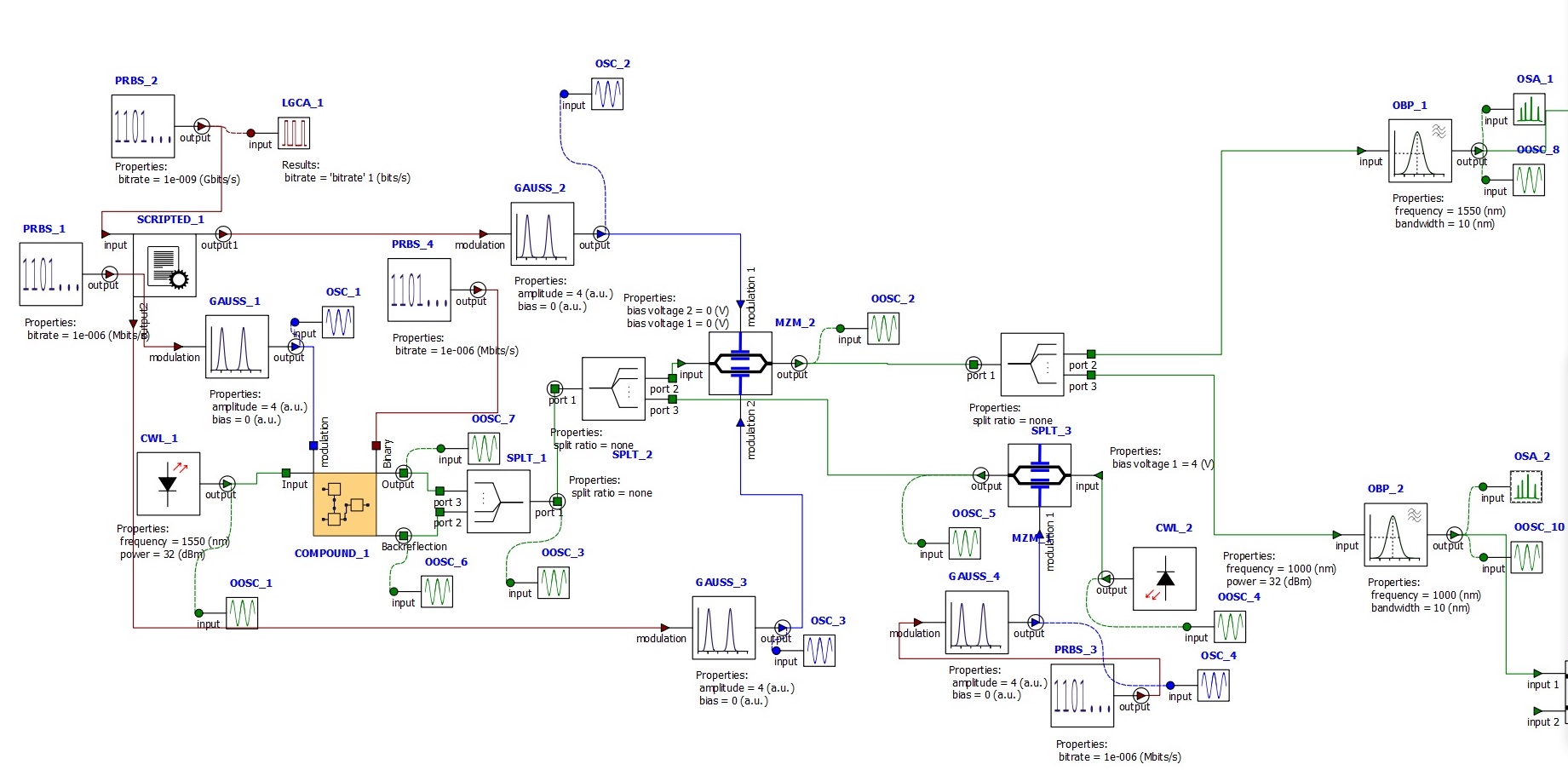}
\caption{Eve attacking Alice's module of ANSYS Interconnect simulation using Trojan-horse attack on DPS protocol}
\end{figure}
\begin{figure}[htp]
\centering
\includegraphics[width=0.25\columnwidth]{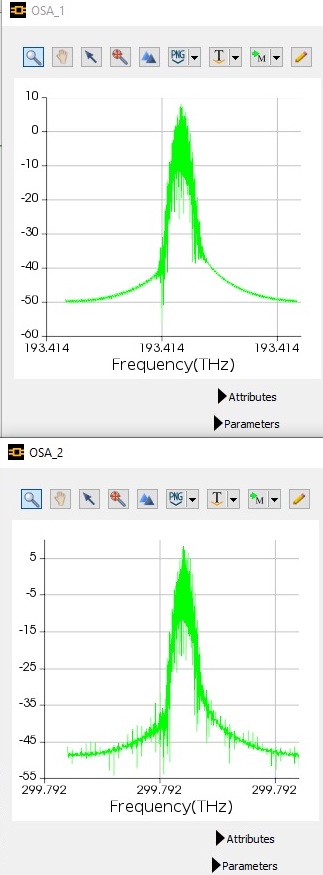}
\caption{Output of Eve attacking Alice's module of ANSYS Interconnect simulation using Trojan-horse attack  on DPS protocol}
\end{figure}
\begin{figure}[htp]
\centering
\includegraphics[width=1\columnwidth]{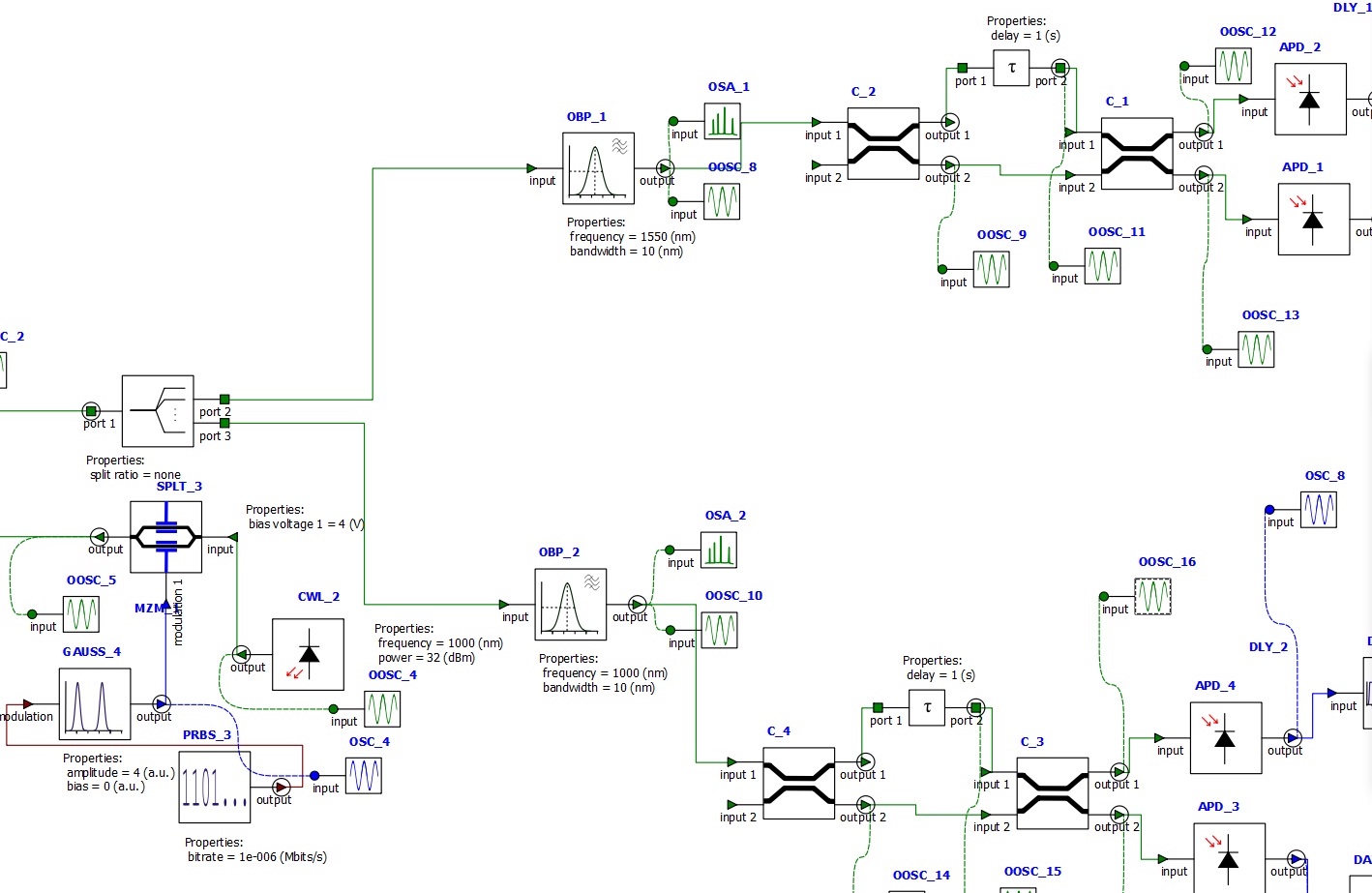}
\caption{Eve and Bob's module of ANSYS Interconnect simulation for Trojan-horse attack on Alice's module on DPS protocol}
\end{figure}
\begin{figure}[htp]
\centering
\includegraphics[width=0.25\columnwidth]{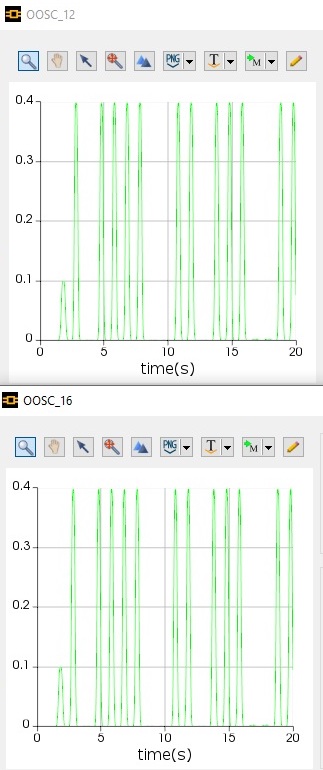}
\caption{Output of Eve and Bob's module of ANSYS Interconnect simulation for Trojan-horse attack on Alice's module on DPS protocol}
\end{figure}
\begin{figure}[htp]
\centering
\includegraphics[width=1\columnwidth]{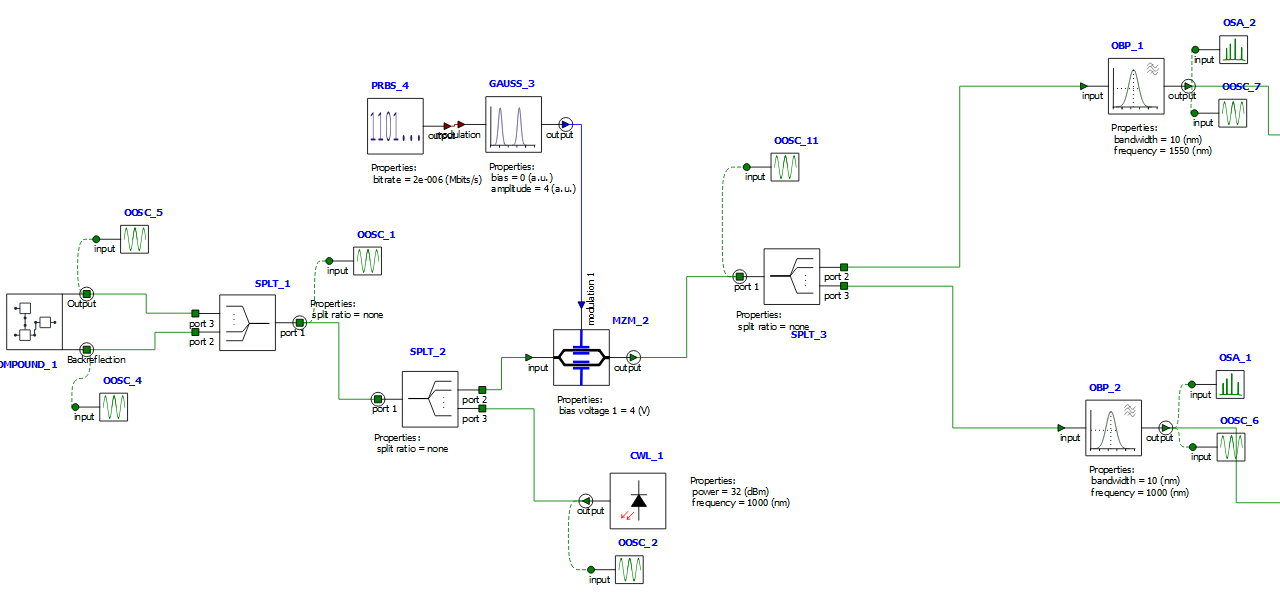}
\caption{Ansys interconnect simulation for Trojan-horse attack on Alices's module on COW protocol}
\end{figure}
\subsection{Countermeasures}
Alice's device could be exposed to intense illumination in a Trojan horse attack. An optical isolator at Alice's entrance can block Eve's light for one-way QKD systems. Installing a second detector at Alice and Bob's device's entrance is a more widely applicable approach. Ideally, this so-called watchdog detector monitors a fraction of all incoming radiation. The QKD function is stopped, and an alert is triggered when the incoming light intensity surpasses a predetermined threshold. A trade-off must be made between the losses brought on and the percentage of light that the watchdog detector is tracking. The idea is to alternate randomly between using the incoming light for monitoring and QKD operation.

\section{Detector blinding attack and pulse illumination attack}
The original detector blinding attack, or continuous wave illumination attack, can be briefly described as Eve exploiting APD between its linear and Geiger modes (non-linear input-output characteristics of SPAD) to eavesdrop information \cite{Lydersen2010}. Eve injects continuous light into the APD in the blinding attack, which lowers the bias voltage and forces the APD back into the linear mode, which is insensitive to a single photon. She then launches a fake state attack. To control Bob's clicks, she first intercepts and measures each state Alice sends, then sends a trigger pulse encoded with the results of her measurement again. A photocurrent monitor is thought to be an effective countermeasure against blinding attacks \cite{Yuan2010}. This countermeasure is based on the logical presumption that a blinding attack will undoubtedly cause a discernible low-frequency photocurrent to emerge in the detector's circuit. The monitor extracts the low-frequency photocurrent as a warning of the blinding attack. The blinding attack is deemed to have begun when the extracted photocurrent crosses a concerning threshold. 

It was recently discovered that this countermeasure using a photocurrent monitor could not prevent the pulse illumination attack \cite{Wu:20}, a mild variant of the detector blinding attack. The pulse illumination attack is more general than the original continuous wave light detector blinding attack. To create a period of complete blinding and detection control, Eve uses a combination of light-shining exploits, such as the ability to switch between the linear mode and the Geiger mode and sophisticated use of the hysteresis current following pulse shining. When intense optical pulses are used to blind an APD, the previously mentioned assumption regarding the photocurrent under a blinding attack is invalidated. A series of blinding pulses progressively introduces a high photocurrent in this attack. It is also possible for this photocurrent to reduce the bias voltage across the APD. As a result, the detector is blinded at that point. The detector remains blinded for a while after the blinding pulses stop because the detector's capacitors cause the photocurrent to drop progressively. Consequently, until the photocurrent becomes relatively weak, the detector remains blinded. Eve can then initiate the fake-state attack to intercept the data by exploiting the blinded period. In theory, a positive correlation exists between the energy of the blinding pulse and the duration of the blinded period. Here, the photocurrent varies with time, in contrast to the continuous high photocurrent caused by the continuous wave illumination attack. The photocurrent monitor ignores most of this current, interpreting it as high-frequency noise. Eve can, therefore, use the pulse illumination attack to intercept the data without the photocurrent monitor noticing.

\subsection{ANSYS simulation of detector blinding attack on DPS protocol}
Bob's implementation in the DPR protocols is ``passive", where he doesn't employ a modulator to add randomization to his detection system \cite{lydersen2011}. For passive implementations like DPS and COW, detector control is more straightforward since Eve does not have to handle the scenarios that arise from Bob choosing a measurement at random (like in the case of the BB84 protocol). Using a duplicate of Bob's setup, Eve measures Alice's states in a detector blinding attack to obtain a detection event. When Eve successfully resends a bright trigger pulse to the detectors operating in the linear mode, Bob's measurement devices experience identical detection events (Alice$\rightarrow$Eve's Bob module$\rightarrow$Eve's faked state generator$\rightarrow$Bob \cite{lydersen2011}). Bob's detection statistics will be identical to those he would receive without Eve because Eve utilizes an exact copy of Bob. As a result, Eve will not be identified by Alice and Bob's data. 

Therefore, the protocol's specifics are pointless because Eve already has an identical copy of Bob's detection results. Eve can listen to the classical channel and apply the same post-processing to her copy of the detection data. Since the detectors may be accessed in the linear mode, the task in a detector blinding attack is to figure out how to cause random detection events in Bob's measuring device.

The APD detectors have two crucial features regardless of how the linear mode is obtained: The minimum trigger pulse strength that always results in a click in any detector is called $P_{always}$, whereas the highest trigger pulse power, $P_{never}$, never causes a click in any detector. $P_{always}<2 P_{never}$ specifies the requirements for a flawless detector blinding attack \cite{Lydersen2010}. Eve's faked state generator (FSG) is essentially a strategy for the DPS and COW protocols \cite{lydersen2011}, which implements coherent pulses with amplitude $P_{always}$ that are brighter than normal in linear mode. And by choosing the phase difference $\phi_k-\phi_{k-1}=(N+1/2)\pi$, 2$N\pi$ and (2$N+1$)$\pi$ for a vacuum event, click in D1 and click in D2, respectively, an arbitrary detector at Bob's side can be activated for pulses in slot $k$ (with $N=\{0,1,2,3\}$), assuming appropriate detection thresholds in DPS protool. 

This FSG can be simulated on DPS protocol initially using varied amplification (AMP\_1 for D1 and AMP\_2 for D2) for the output of D1 (SPAD 0 phase diff) and D2 (SPAD2 Pi phase diff) detectors of Eve's FSG module with Bob's setup as seen in Fig. 17. The output of this for a particular instance of detection for Eve is shown in Fig. 18, where in our case amplitude zero indicates the no detection, amplitude one indicates bit 0 (Click in D1 detector) and amplitude two indicates bit 1 (Click in D2 detector). Further, based on this output, we select the phase of the pulse sent to Bob by Eve, where $N\pi/2$ is the phase decided for $N=0,1,2,3$. For this particular instance of reading $\{0, 1, 2, 0, 1, 2, 2, 0, 2, 2, 0, 2, 0, 0, 0\}$ in Fig. 18 by Eve, we get the phase encoding for Eve's FSG as $\{0, 0, 2, 1, 1, 3, 1, 2, 0, 2, 1, 3, 2, 1, 2\}$. We use this phase information to encode the pulses and observe the output of Bob's module in Fig. 19 using a setup similar to Fig. 1, where Eve becomes Alice. As can be seen, $P_{never}$ pulses are in the range of 0.2 amplitude, and $P_{always}$ pulses are in the range of 0.4 amplitude at the detectors.

\begin{figure}[htp]
\centering
\includegraphics[width=1\columnwidth]{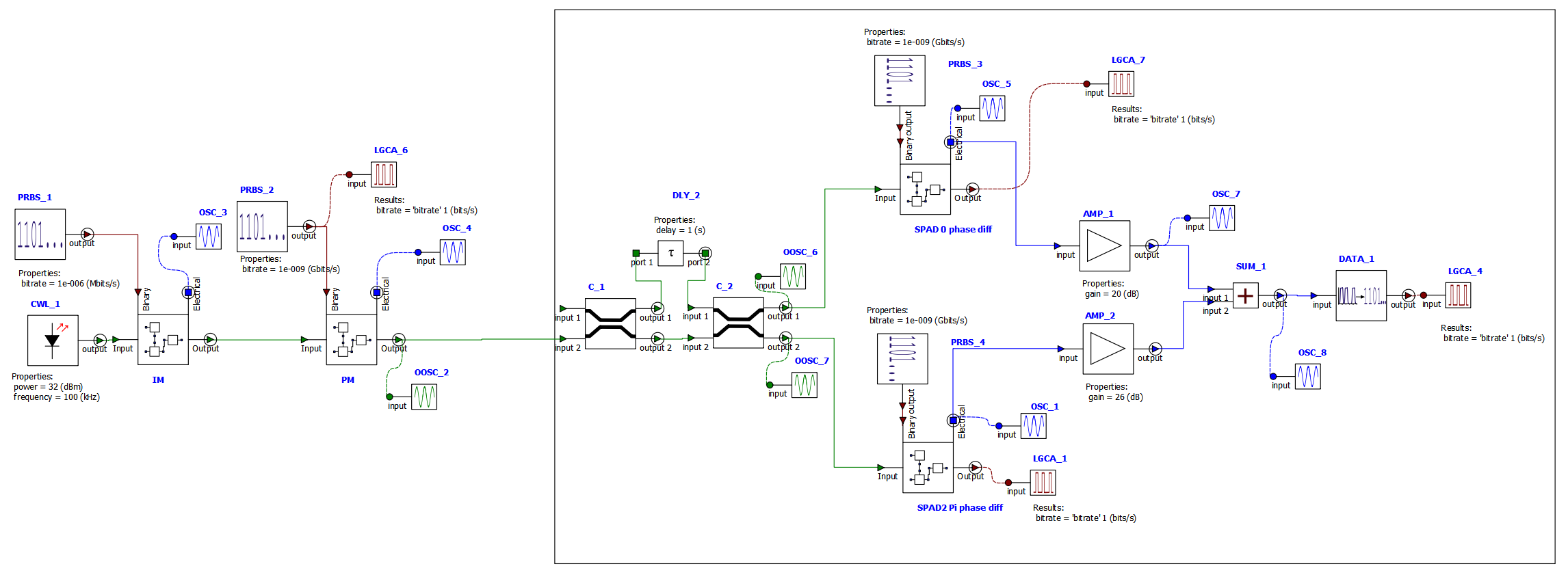}
\caption{ANSYS Interconnect simulation for detector blinding attack on DPS protocol. We reproduce Bob's setup and detection part of Eve's FSG which is highlighted in the rectangular selection.}
\end{figure}
\begin{figure}[htp]
\centering
\includegraphics[width=0.5\columnwidth]{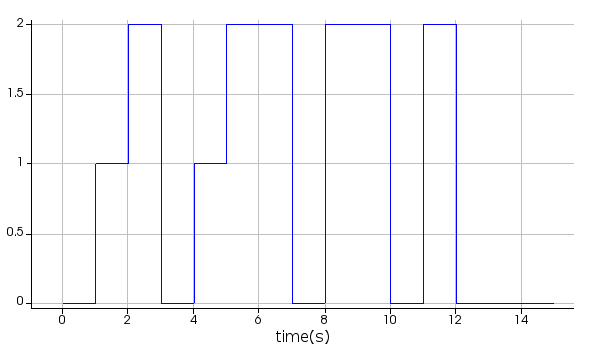}
\caption{Initial Eve's detection which is used to select $\phi$ of phase modulator in Eve's FSG. Here 0 indicates no detection, 1 indicates bit 0 and 2 indicates 1 detection.}
\end{figure}
\begin{figure}[htp]
\centering
\begin{subfigure}{0.45\linewidth}
\includegraphics[width=\linewidth]{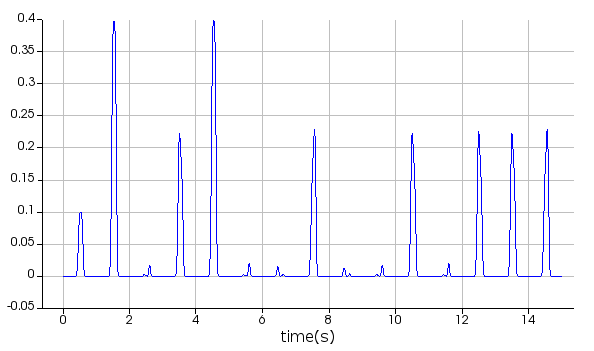}
\caption{Power at D1 of Bob}
\end{subfigure}
\begin{subfigure}{0.45\linewidth}
\includegraphics[width=\linewidth]{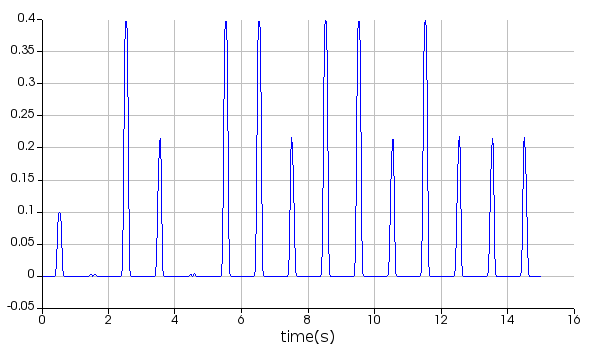}
\caption{Power at D2 of Bob}
\end{subfigure}
\caption{Power output Bob's module based on Eve's detection in Fig. 18.}
\end{figure}
\subsection{ANSYS simulation of detector blinding attack on COW protocol}
Eve must determine distinct trigger pulse thresholds for the monitoring detectors $D_{M1}$ (APD\_2) and $D_{M2}$ (APD\_3), as well as the data detector $D_{B}$ (APD\_1), to perform eavesdropping against COW as can be seen in Fig. 2. The data detector's thresholds are $P_{always,B}$ and $P_{never,B}$, while the monitoring detectors' thresholds are $P_{always,M}$ and $P_{never,M}$.

The monitoring detectors have the same setup as DPS, and they can be regulated as explained in the previous section. However, the amplitude of the pulse train needs to be increased to $P_{always,M}/(1-t_B)$ since only $(1-t_B)$ of the trigger pulse power enters the interferometer of the monitoring detectors ($t_B=0.9$ in COW simulation from Fig.2). For flawless control, the data detector isn't activated by light entering the other arm. For this to work, \begin{equation}
    \frac{t_B}{1-t_B}P_{always,M}<P_{never,B}.
\end{equation} The trigger pulse's amplitude can be increased to $P_{always,B}/t_B$ in order to activate the data detector. But it's imperative that the monitoring detectors not go off at that point. The phase difference is adjusted to $\pi/2$ to minimise the illumination on the monitoring detectors. The threshold requirement is given by \begin{equation}
    \frac{1-t_B}{t_B}P_{always,B}<2P_{never,M},
\end{equation} where factor 2 indicates that the illumination is split between the two monitoring detectors. The data detector's thresholds need to be significantly higher than the monitoring detectors when $t_B$ is near 1, which we avoid for convenience and consider $t_B=0.5$ (similar to Ref.\cite{lydersen2011}).

\begin{figure}[htp]
\centering
\includegraphics[width=1\columnwidth]{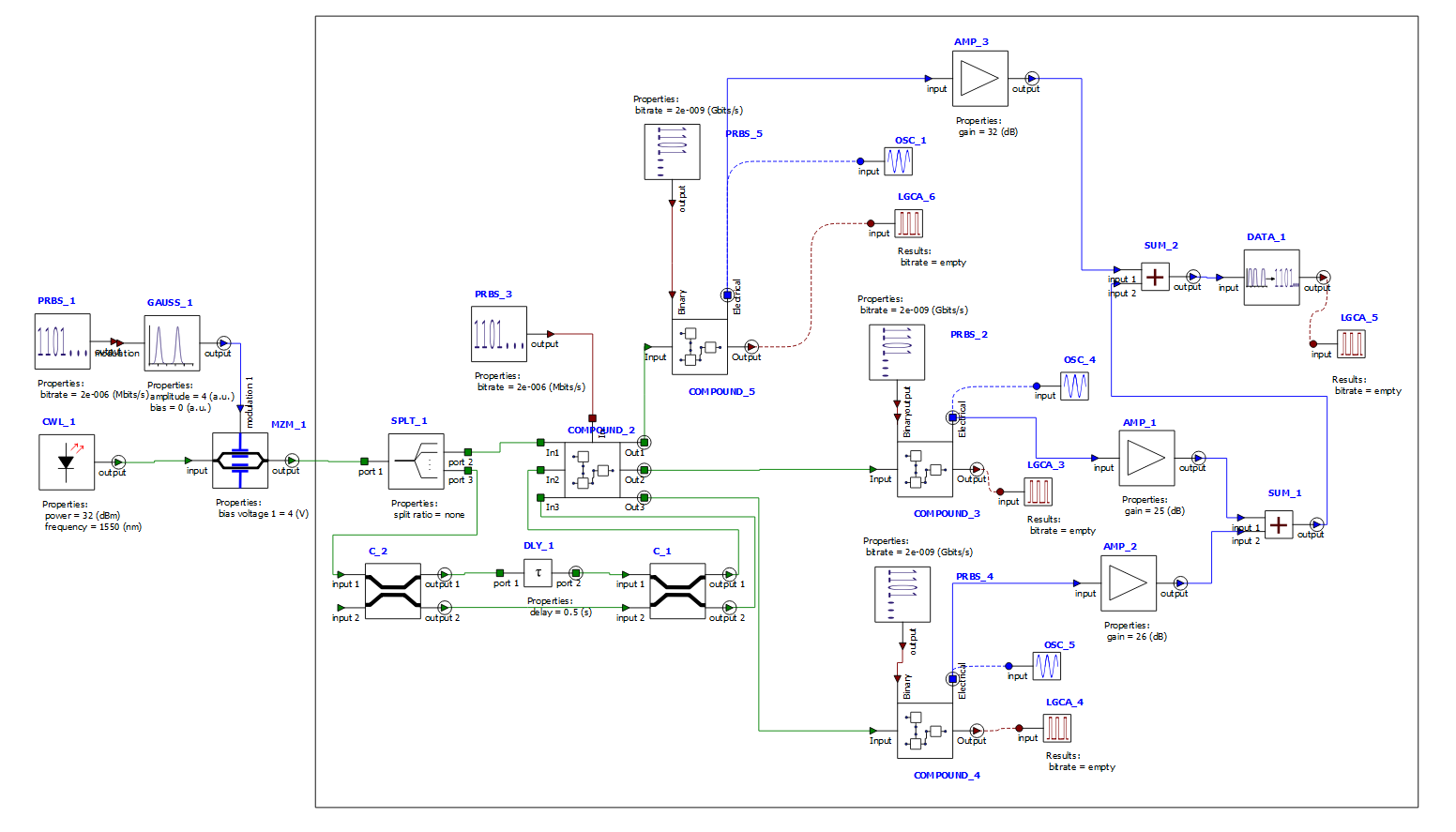}
\caption{ANSYS Interconnect simulation for detector blinding attack on COW protocol. We reproduce Bob's setup and detection part of Eve's FSG which is highlighted in the rectangular selection.}
\end{figure}
This FSG can be simulated on COW protocol initially using varied amplification (AMP\_1 for $D_{M1}$, AMP\_2 for $D_{M2}$ and AMP\_3 for $D_{B}$) for the output of detectors of Eve's module of Bob as seen in Fig. 20. 
\begin{figure}[htp]
\centering
\includegraphics[width=0.5\columnwidth]{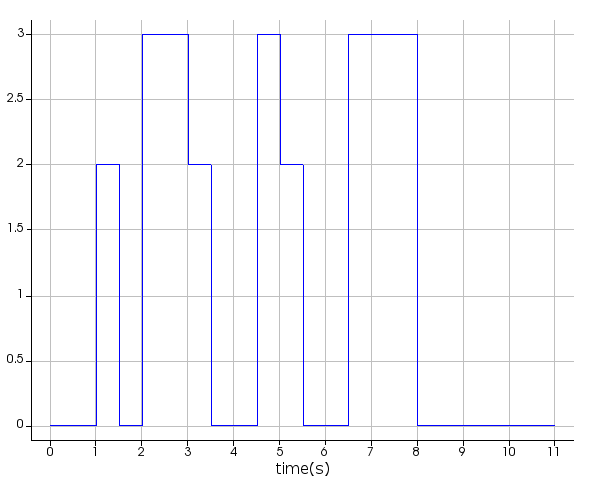}
\caption{Initial Eve's detection which is used to select $\phi$ and amplitude modulation. Here 0 indicates no detection, 1 indicates detection at $D_{M2}$, 2 indicates detection at $D_{M1}$ and 3 indicates detection at $D_{B}$.}
\end{figure}

The output of this for a particular instance of detection for Eve is shown in Fig. 21, where amplitude zero indicates no detection, one indicates detection at $D_{M2}$, two indicates detection at $D_{M1}$, and three indicates detection at $D_{B}$. Further, based on this output, we create Eve's FSG output to select the phase and amplitude of the pulse sent to Bob to regulate the monitoring detectors and detection line, almost exactly replicating Alice's optical approach. A coherent pulse train with amplitude $P_{always,M}/(1-t_B)$ is released by Eve. We increase the amplitude to $P_{always,M}/(t_B)$ to trigger the data detector as shown in Fig. 22. To activate one of the monitoring detectors in slot $k$, we choose the phase difference $\phi_k-\phi_{k-1}=(N+1/2)\pi$, 2$N\pi$ and (2$N+1$)$\pi$ for a vacuum event, click in APD\_2 and click in APD\_1, respectively (with $N=\{0,1,2,3\}$) similar to DPS detectors control in previous section. It is simple to create simultaneous clicks in one of the monitoring detectors and the data detector because they are regulated independently. For instance, in Fig. 21, the output on Bob's side is shown in Fig. 23.

\begin{figure}[htp]
\centering
\includegraphics[width=1\columnwidth]{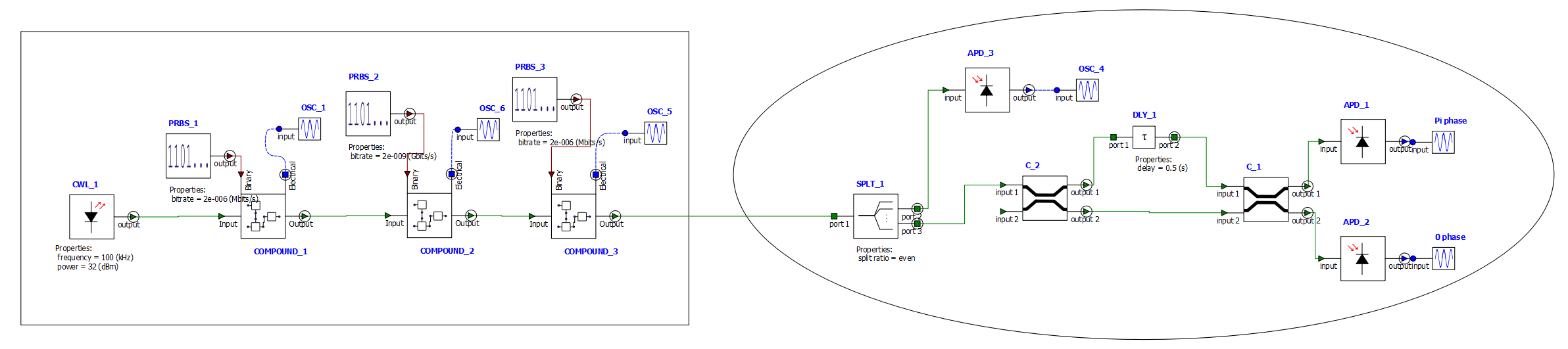}
\caption{ANSYS Interconnect simulation for detector blinding attack on COW protocol with Eve's FSG output which is highlighted in the rectangular selection and Bob's setup is shown in oval selection.}
\end{figure}
\begin{figure}[htp]
\centering
\begin{subfigure}{0.45\linewidth}
\includegraphics[width=\linewidth]{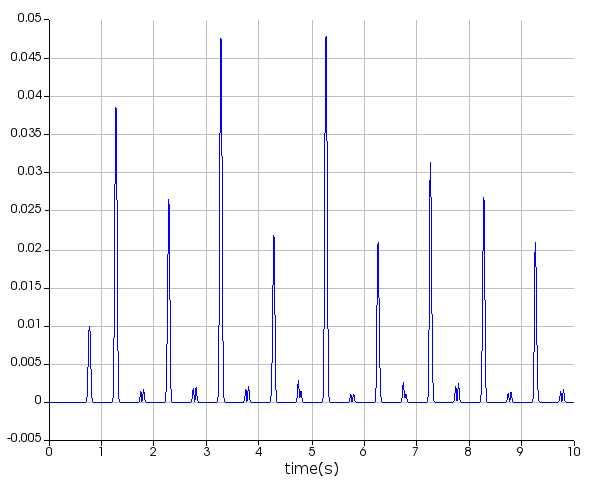}
\caption{Power at $D_{M1}$ (APD\_1) of Bob}
\end{subfigure}
\begin{subfigure}{0.45\linewidth}
\includegraphics[width=\linewidth]{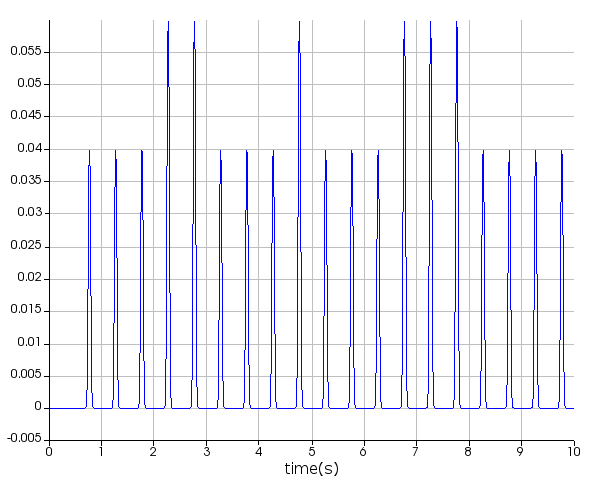}
\caption{Power at $D_B$ (APD\_3) of Bob}
\end{subfigure}
\caption{Power output Bob's module based on Eve's detection in Fig. 21.}
\end{figure}
\subsection{Countermeasures}
The photocurrent monitor's alarm threshold must be lowered to detect the pulse illumination attack \cite{Wu:20}. This method might introduce a lot of false alarms, but Eve can break this defence by enlarging the intervals, which attenuates the reported photocurrent. Enhancing the photocurrent monitor's filter's stop band can uncover additional proof of the pulse illumination attack. Enhancing the stop band to an extreme degree will also cause the monitor to malfunction and generate a lot of false alarms from noises. Removing the filter to reveal the pulse illumination attack fully will undoubtedly set off the alarm, but it's also possible for the detector to frequently set off false alarms when operating normally.
\section{Conclusions}

We simulated hacking attempts like backflash, Trojan-horse, and detector-control attacks on differential-phase-shift and coherent-one-way quantum key distribution protocols. These simulations can further be improved to match realistic experimental parameters and quantify the information lost by the hacking attempt in the presence of an eavesdropper. 

Attack ratings were recently assessed for all the possible attacks in Ref. \cite{IAQKD}, indicating how hard it is to carry out a suggested attack correctly and effectively. It is determined by factors such as elapsed time, expertise of eavesdropper, target of evaluation knowledge, equipment, and window of opportunity. These elements are taken from the most recent version of the common criteria methodology, a trustworthy method for estimating attack ratings on cryptographic systems. The attacks considered in this work are rated between "basic" and "high" based on all their various factors.

It is widely held that flaws and vulnerabilities are inherent to any technology still in development, and practical QKD is no exception. 

\bibliographystyle{ieeetr}
\bibliography{sample.bib}

\begin{thebibliography}{10}

\bibitem{PhysRevLett.67.661}
A.~K. Ekert, ``Quantum cryptography based on bell's theorem,'' {\em Phys. Rev. Lett.}, vol.~67, pp.~661--663, Aug 1991.

\bibitem{RevModPhys.81.1301}
V.~Scarani, H.~Bechmann-Pasquinucci, N.~J. Cerf, M.~Du\ifmmode~\check{s}\else \v{s}\fi{}ek, N.~L\"utkenhaus, and M.~Peev, ``The security of practical quantum key distribution,'' {\em Rev. Mod. Phys.}, vol.~81, pp.~1301--1350, Sep 2009.

\bibitem{Lo2014}
H.-K. Lo, M.~Curty, and K.~Tamaki, ``Secure quantum key distribution,'' {\em Nature Photonics}, vol.~8, pp.~595--604, Aug 2014.

\bibitem{RevModPhys.92.025002}
F.~Xu, X.~Ma, Q.~Zhang, H.-K. Lo, and J.-W. Pan, ``Secure quantum key distribution with realistic devices,'' {\em Rev. Mod. Phys.}, vol.~92, p.~025002, May 2020.

\bibitem{Pirandola:20}
S.~Pirandola, U.~L. Andersen, L.~Banchi, M.~Berta, D.~Bunandar, R.~Colbeck, D.~Englund, T.~Gehring, C.~Lupo, C.~Ottaviani, J.~L. Pereira, M.~Razavi, J.~S. Shaari, M.~Tomamichel, V.~C. Usenko, G.~Vallone, P.~Villoresi, and P.~Wallden, ``Advances in quantum cryptography,'' {\em Adv. Opt. Photon.}, vol.~12, pp.~1012--1236, Dec 2020.

\bibitem{BENNETT20147}
C.~H. Bennett and G.~Brassard, ``Quantum cryptography: Public key distribution and coin tossing,'' {\em Theoretical Computer Science}, vol.~560, pp.~7--11, 2014.
\newblock Theoretical Aspects of Quantum Cryptography – celebrating 30 years of BB84.

\bibitem{PhysRevLett.89.037902}
K.~Inoue, E.~Waks, and Y.~Yamamoto, ``Differential phase shift quantum key distribution,'' {\em Phys. Rev. Lett.}, vol.~89, p.~037902, Jun 2002.

\bibitem{Damien}
D.~Stucki, N.~Brunner, N.~Gisin, V.~Scarani, and H.~Zbinden, ``{Fast and simple one-way quantum key distribution},'' {\em Applied Physics Letters}, vol.~87, p.~194108, 11 2005.

\bibitem{Stucki:09}
D.~Stucki, C.~Barreiro, S.~Fasel, J.-D. Gautier, O.~Gay, N.~Gisin, R.~Thew, Y.~Thoma, P.~Trinkler, F.~Vannel, and H.~Zbinden, ``Continuous high speed coherent one-way quantum key distribution,'' {\em Opt. Express}, vol.~17, pp.~13326--13334, Aug 2009.

\bibitem{Stucki_2009}
D.~Stucki, N.~Walenta, F.~Vannel, R.~T. Thew, N.~Gisin, H.~Zbinden, S.~Gray, C.~R. Towery, and S.~Ten, ``High rate, long-distance quantum key distribution over 250 km of ultra low loss fibres,'' {\em New Journal of Physics}, vol.~11, p.~075003, jul 2009.

\bibitem{Walenta_2014}
N.~Walenta, A.~Burg, D.~Caselunghe, J.~Constantin, N.~Gisin, O.~Guinnard, R.~Houlmann, P.~Junod, B.~Korzh, N.~Kulesza, M.~Legré, C.~W. Lim, T.~Lunghi, L.~Monat, C.~Portmann, M.~Soucarros, R.~T. Thew, P.~Trinkler, G.~Trolliet, F.~Vannel, and H.~Zbinden, ``A fast and versatile quantum key distribution system with hardware key distillation and wavelength multiplexing,'' {\em New Journal of Physics}, vol.~16, p.~013047, jan 2014.

\bibitem{Korzh2015}
B.~Korzh, C.~C.~W. Lim, R.~Houlmann, N.~Gisin, M.~J. Li, D.~Nolan, B.~Sanguinetti, R.~Thew, and H.~Zbinden, ``Provably secure and practical quantum key distribution over 307 km of optical fibre,'' {\em Nature Photonics}, vol.~9, pp.~163--168, Mar 2015.

\bibitem{Sibson2017}
P.~Sibson, C.~Erven, M.~Godfrey, S.~Miki, T.~Yamashita, M.~Fujiwara, M.~Sasaki, H.~Terai, M.~G. Tanner, C.~M. Natarajan, R.~H. Hadfield, J.~L. O'Brien, and M.~G. Thompson, ``Chip-based quantum key distribution,'' {\em Nature Communications}, vol.~8, p.~13984, Feb 2017.

\bibitem{Sibson:17}
P.~Sibson, J.~E. Kennard, S.~Stanisic, C.~Erven, J.~L. O'Brien, and M.~G. Thompson, ``Integrated silicon photonics for high-speed quantum key distribution,'' {\em Optica}, vol.~4, pp.~172--177, Feb 2017.

\bibitem{roberts2017}
G.~L. Roberts, M.~Lucamarini, J.~F. Dynes, S.~J. Savory, Z.~L. Yuan, and A.~J. Shields, ``Modulator-free coherent-one-way quantum key distribution,'' {\em Laser \& Photonics Reviews}, vol.~11, no.~4, p.~1700067, 2017.

\bibitem{Dai:20}
J.~Dai, L.~Zhang, X.~Fu, X.~Zheng, and L.~Yang, ``Pass-block architecture for distributed-phase-reference quantum key distribution using silicon photonics,'' {\em Opt. Lett.}, vol.~45, pp.~2014--2017, Apr 2020.

\bibitem{Takesue_2005}
H.~Takesue, E.~Diamanti, T.~Honjo, C.~Langrock, M.~M. Fejer, K.~Inoue, and Y.~Yamamoto, ``Differential phase shift quantum key distribution experiment over 105 km fibre,'' {\em New Journal of Physics}, vol.~7, p.~232, nov 2005.

\bibitem{Diamanti:06}
E.~Diamanti, H.~Takesue, C.~Langrock, M.~M. Fejer, and Y.~Yamamoto, ``100 km differential phase shift quantum key distribution experiment with low jitter up-conversion detectors,'' {\em Opt. Express}, vol.~14, pp.~13073--13082, Dec 2006.

\bibitem{Takesue2007}
H.~Takesue, S.~W. Nam, Q.~Zhang, R.~H. Hadfield, T.~Honjo, K.~Tamaki, and Y.~Yamamoto, ``Quantum key distribution over a 40-db channel loss using superconducting single-photon detectors,'' {\em Nature Photonics}, vol.~1, pp.~343--348, Jun 2007.

\bibitem{Sasaki:11}
M.~Sasaki, M.~Fujiwara, H.~Ishizuka, W.~Klaus, K.~Wakui, M.~Takeoka, S.~Miki, T.~Yamashita, Z.~Wang, A.~Tanaka, K.~Yoshino, Y.~Nambu, S.~Takahashi, A.~Tajima, A.~Tomita, T.~Domeki, T.~Hasegawa, Y.~Sakai, H.~Kobayashi, T.~Asai, K.~Shimizu, T.~Tokura, T.~Tsurumaru, M.~Matsui, T.~Honjo, K.~Tamaki, H.~Takesue, Y.~Tokura, J.~F. Dynes, A.~R. Dixon, A.~W. Sharpe, Z.~L. Yuan, A.~J. Shields, S.~Uchikoga, M.~Legr\'{e}, S.~Robyr, P.~Trinkler, L.~Monat, J.-B. Page, G.~Ribordy, A.~Poppe, A.~Allacher, O.~Maurhart, T.~L\"{a}nger, M.~Peev, and A.~Zeilinger, ``Field test of quantum key distribution in the tokyo qkd network,'' {\em Opt. Express}, vol.~19, pp.~10387--10409, May 2011.

\bibitem{IAQKD}
B.~federal office for~information security, ``Implementation attacks against qkd systems,'' {\em https://www.bsi.bund.de/SharedDocs/Downloads/EN/BSI/Publications/Studies/QKD-Systems/QKD-Systems.pdf}, Dec 2023.

\bibitem{Shuang2008}
S.~{Zhao} and H.~{De Raedt}, ``{Event-by-Event Simulation of Quantum Cryptography Protocols},'' {\em Journal of Computational and Theoretical Nanoscience}, vol.~5, pp.~490--504, Apr. 2008.

\bibitem{Buhari2012}
A.~Buhari, Z.~A. Zukarnain, S.~K. Subramaniam, H.~Zainuddin, and S.~Saharudin, ``An efficient modeling and simulation of quantum key distribution protocols using optisystem™,'' in {\em 2012 IEEE Symposium on Industrial Electronics and Applications}, pp.~84--89, 2012.

\bibitem{mailloux2015}
L.~O. Mailloux, J.~D. Morris, M.~R. Grimaila, D.~D. Hodson, D.~R. Jacques, J.~M. Colombi, C.~V. Mclaughlin, and J.~A. Holes, ``A modeling framework for studying quantum key distribution system implementation nonidealities,'' {\em IEEE Access}, vol.~3, pp.~110--130, 2015.

\bibitem{mailloux2016}
L.~O. Mailloux, D.~D. Hodson, M.~R. Grimaila, R.~D. Engle, C.~V. Mclaughlin, and G.~B. Baumgartner, ``Using modeling and simulation to study photon number splitting attacks,'' {\em IEEE Access}, vol.~4, pp.~2188--2197, 2016.

\bibitem{Coles2016}
P.~J. Coles, E.~M. Metodiev, and N.~L{\"u}tkenhaus, ``Numerical approach for unstructured quantum key distribution,'' {\em Nature Communications}, vol.~7, p.~11712, May 2016.

\bibitem{PhysRevApplied.14.024036}
R.~Chatterjee, K.~Joarder, S.~Chatterjee, B.~C. Sanders, and U.~Sinha, ``qkdsim, a simulation toolkit for quantum key distribution including imperfections: Performance analysis and demonstration of the b92 protocol using heralded photons,'' {\em Phys. Rev. Appl.}, vol.~14, p.~024036, Aug 2020.

\bibitem{Fan-Yuan2020}
G.-J. Fan-Yuan, W.~Chen, F.-Y. Lu, Z.-Q. Yin, S.~Wang, G.-C. Guo, and Z.-F. Han, ``A universal simulating framework for quantum key distribution systems,'' {\em Science China Information Sciences}, vol.~63, p.~180504, Jul 2020.

\bibitem{Anuj2022}
A.~Sethia and A.~Banerjee, ``A matlab-based modelling and simulation package for dps-qkd,'' {\em Journal of Modern Optics}, vol.~69, no.~7, pp.~392--402, 2022.

\bibitem{Jain2015}
N.~Jain, B.~Stiller, I.~Khan, V.~Makarov, C.~Marquardt, and G.~Leuchs, ``Risk analysis of trojan-horse attacks on practical quantum key distribution systems,'' {\em IEEE Journal of Selected Topics in Quantum Electronics}, vol.~21, no.~3, pp.~168--177, 2015.

\bibitem{Pinheiro:18}
P.~V.~P. Pinheiro, P.~Chaiwongkhot, S.~Sajeed, R.~T. Horn, J.-P. Bourgoin, T.~Jennewein, N.~L\"{u}tkenhaus, and V.~Makarov, ``Eavesdropping and countermeasures for backflash side channel in quantum cryptography,'' {\em Opt. Express}, vol.~26, pp.~21020--21032, Aug 2018.

\bibitem{Meda2017}
A.~Meda, I.~P. Degiovanni, A.~Tosi, Z.~Yuan, G.~Brida, and M.~Genovese, ``Quantifying backflash radiation to prevent zero-error attacks in quantum key distribution,'' {\em Light: Science {\&} Applications}, vol.~6, pp.~e16261--e16261, Jun 2017.

\bibitem{jmo2001}
V.~M. Artem~Vakhitov and D.~R. Hjelme, ``Large pulse attack as a method of conventional optical eavesdropping in quantum cryptography,'' {\em Journal of Modern Optics}, vol.~48, no.~13, pp.~2023--2038, 2001.

\bibitem{th2006}
N.~Gisin, S.~Fasel, B.~Kraus, H.~Zbinden, and G.~Ribordy, ``Trojan-horse attacks on quantum-key-distribution systems,'' {\em Phys. Rev. A}, vol.~73, p.~022320, Feb 2006.

\bibitem{Jain_2014}
N.~Jain, E.~Anisimova, I.~Khan, V.~Makarov, C.~Marquardt, and G.~Leuchs, ``Trojan-horse attacks threaten the security of practical quantum cryptography,'' {\em New Journal of Physics}, vol.~16, p.~123030, dec 2014.

\bibitem{Sajeed2017}
S.~Sajeed, C.~Minshull, N.~Jain, and V.~Makarov, ``Invisible trojan-horse attack,'' {\em Scientific Reports}, vol.~7, p.~8403, Aug 2017.

\bibitem{Lydersen2010}
L.~Lydersen, C.~Wiechers, C.~Wittmann, D.~Elser, J.~Skaar, and V.~Makarov, ``Hacking commercial quantum cryptography systems by tailored bright illumination,'' {\em Nature Photonics}, vol.~4, pp.~686--689, Oct 2010.

\bibitem{Wu:20}
Z.~Wu, A.~Huang, H.~Chen, S.-H. Sun, J.~Ding, X.~Qiang, X.~Fu, P.~Xu, and J.~Wu, ``Hacking single-photon avalanche detectors in quantum key distribution via pulse illumination,'' {\em Opt. Express}, vol.~28, pp.~25574--25590, Aug 2020.

\bibitem{PhysRev.100.700}
R.~Newman, ``Visible light from a silicon $p\ensuremath{-}n$ junction,'' {\em Phys. Rev.}, vol.~100, pp.~700--703, Oct 1955.

\bibitem{6570518}
F.~Acerbi, A.~Tosi, and F.~Zappa, ``Avalanche current waveform estimated from electroluminescence in ingaas/inp spads,'' {\em IEEE Photonics Technology Letters}, vol.~25, no.~18, pp.~1778--1780, 2013.

\bibitem{Yuan2010}
Z.~L. Yuan, J.~F. Dynes, and A.~J. Shields, ``Avoiding the blinding attack in qkd,'' {\em Nature Photonics}, vol.~4, pp.~800--801, Dec 2010.

\bibitem{lydersen2011}
J.~S. L.~Lydersen and V.~Makarov, ``Tailored bright illumination attack on distributed-phase-reference protocols,'' {\em Journal of Modern Optics}, vol.~58, no.~8, pp.~680--685, 2011.

\end{thebibliography}
\end{document}